\RequirePackage{amsmath}
\documentclass[epj,nopacs]{svjour}

\usepackage{graphicx}
\usepackage{mathtools}
\usepackage{axodraw2}
\usepackage{amsmath}
\usepackage{amssymb}
\usepackage{breqn}
\usepackage{url}
\usepackage{bm}
\usepackage{slashed}
\usepackage{color}

\begin{document} 

\title{Naturalising the Third Family Hypercharge Model for Neutral Current $B-$Anomalies}   


\author{B.C. Allanach and Joe Davighi}
\institute{DAMTP, University of Cambridge, Wilberforce Road, Cambridge, CB3 0WA, UK}

\abstract{We consider a deformation of the Third Family Hypercharge Model,
  which arguably makes the model more natural. Additional non-zero charges of the
  spontaneously broken,
  family-dependent $U(1)_F$ gauge symmetry are assigned 
  to the second family leptons, and the third family leptons' charges are deformed away
  from their hypercharges in such a way that the $U(1)_F$ gauge symmetry
  remains anomaly-free. Second family $U(1)_F$ lepton charges allow a
  $Z^\prime$ coupling to muons without having to assume large charged lepton
  mixing, which risks violating tight lepton flavour violation bounds.
  In this deformed version, only the bottom and top Yukawa couplings are
  generated at the renormalisable level, 
  whereas the tauon Yukawa coupling is absent. The $Z^\prime$  mediates   a 
  beyond the Standard Model contribution
  to an effective $(\bar b s) (\bar \mu
  \mu)$ vertex in the combination $C_9=-9C_{10}$ and is able to fit the apparent discrepancy between Standard Model
  predictions in flavour changing neutral-current $B-$meson decays and their
  measurements,
  whilst simultaneously avoiding current constraints from direct $Z^\prime$
  searches and other measurements, when $1.2 \text{~TeV} < M_{Z^\prime} < 12.5
  \text{~TeV}$.  
}

\authorrunning{B.C. Allanach \and J. Davighi}
\titlerunning{Naturalising the TFHM}

\maketitle
\flushbottom

\section{Introduction}
\sloppy Various measurements of $B$ meson decays are currently in tension with
Standard Model predictions. 
For instance, the ratio of branching ratios 
$$R_{K^{(\ast)}} \equiv BR(B \rightarrow K^{(\ast)} \mu^+ \mu^-) / BR(B
\rightarrow K^{(\ast)} e^+ e^-)$$ is predicted to be 1.00 in the Standard Model 
(SM),
for lepton invariant mass squared bin $m_{ll}^2 \in [1.1,6]$ GeV$^2$. In this bin,
LHCb measurements~\cite{Aaij:2017vbb,CERN-EP-2019-043} imply $R_K=0.846^{+0.060}_{-0.054}{}^{+0.016}_{-0.014}$ and $R_{K^{\ast}}=0.69^{+0.11}_{-0.07}\pm0.05$. The branching
ratio $B_s \rightarrow \mu^+ \mu^-$~\cite{Aaboud:2018mst,Chatrchyan:2013bka,CMS:2014xfa,Aaij:2017vad} is also measured to be lower than the SM
prediction, which is accurate at the percent level. Angular
distributions in the $B \rightarrow K^{(\ast)} \mu^+ \mu^-$ decays have~\cite{Aaij:2013qta,Aaij:2015oid,ATLAS-CONF-2017-023,CMS-PAS-BPH-15-008} a higher level of disagreement with SM predictions~\cite{Khachatryan:2015isa,Bobeth:2017vxj}, although 
theoretical uncertainties in the SM predictions are larger, at the ten(s)
percent level. There are 
several other indications of disagreements between SM 
predictions and measurements involving the $(\bar s b) (\bar \mu \mu)$
effective coupling. Henceforth, we shall collectively call these disagreements the Neutral Current
$B-$Anomalies (NCBAs). 

We begin with the effective Lagrangian pertinent to the NCBAs\footnote{Fermion
fields are written in the mass eigenbasis unless they are primed, in which
case they are in the weak eigenbasis.}
\begin{eqnarray} \label{bsmm}
  {\mathcal L}_{bs\mu\mu} =
  -\frac{C_L}{(36 \text{~TeV})^2} (\overline{s_L} \gamma_\rho b_L) 
  (\overline{\mu_L} \gamma^\rho \mu_L) + \\
  -\frac{C_R}{(36 \text{~TeV})^2} (\overline{s_L} \gamma_\rho b_L) 
  (\overline{\mu_R} \gamma^\rho \mu_R), \nonumber 
\end{eqnarray}
where we are currently neglecting a contribution from right-handed quarks
because there is no strong evidence in its favour from the data.
The dimensionful denominator in front of each effective coupling is equal to
$4 \pi v^2/(V_{tb} V_{ts}^\ast \alpha)$, where $v=174$ GeV is the SM Higgs
vacuum 
expectation value (VEV), $\alpha$ is the fine structure constant and
$V_{tb}$ and $V_{ts}$ are Cabbibo-Kobayashi-Maskawa (CKM) matrix elements\footnote{$V_{ts}$ has a
  negligible imaginary component, which we neglect.}. 
The SM
contributes $C_L^{SM}=8.64$ and $C_R^{SM}=-0.18$~\cite{DAmico:2017mtc}, the
dominant contributions to each being from one-loop Feynman
diagrams involving $W$ bosons. 

As discussed above,
current data strongly favour a beyond the SM (BSM) contribution to $C_L$ and 
possibly $C_R$~\cite{Alguero:2019ptt,Alok:2019ufo,Ciuchini:2019usw,Aebischer:2019mlg,Kowalska:2019ley,Arbey:2019duh}. One possibility to generate this at tree-level is
by a heavy $Z^\prime$ vector boson that has flavour
non-universal interactions including
\begin{equation}
  {\mathcal L}_{bs\mu\mu}^{Z^\prime}=  -g_{\mu_L}
      \overline{\mu_L} \slashed{Z}^\prime
    \mu_L - g_{\mu_R} \overline{\mu_R} \slashed{Z}^\prime \mu_R -
    g_{sb} \left( \overline{s_L}\slashed{Z}^\prime b_L +H.c.\right).
    \label{NCBAZ}
\end{equation}
Once the $Z^\prime$ is integrated out of the theory (such that the
appropriate theory is the SM effective field theory SMEFT), one obtains the
operators
\begin{eqnarray} \label{eq:SMEFT}
  {\mathcal L}_{bs\mu\mu}^{\mathrm{SMEFT}}= -\frac{g_{sb} g_{\mu_L}}{M_{Z^\prime}^2}
  (\overline{s_L} \gamma^\rho b_L) (\overline{\mu_L} \gamma_\rho \mu_L) + \\
    -\frac{g_{sb} g_{\mu_R}}{M_{Z^\prime}^2}
      (\overline{s_L} \gamma^\rho b_L) (\overline{\mu_R} \gamma_\rho \mu_R). \nonumber
  \end{eqnarray}
Matching Eq.~\ref{eq:SMEFT} with Eq.~\ref{bsmm} identifies
$C_{L,R}=g_{sb} g_{\mu_{L,R}} (36 \text{~TeV}/M_{Z^\prime})^2$.

Many models based on spontaneously broken flavour-dependent gauged $U(1)$ symmetries~\cite{Ellis:2017nrp,Allanach:2018vjg} have been
proposed from which such $Z^\prime$s may result, for example from $L_\mu-L_\tau$ and related groups~\cite{Gauld:2013qba,Buras:2013dea,Buras:2013qja,Altmannshofer:2014cfa,Buras:2014yna,Crivellin:2015mga,Crivellin:2015lwa,Sierra:2015fma,Crivellin:2015era,Celis:2015ara,Greljo:2015mma,Altmannshofer:2015mqa,Allanach:2015gkd,Falkowski:2015zwa,Chiang:2016qov,Becirevic:2016zri,Boucenna:2016wpr,Boucenna:2016qad,Ko:2017lzd,Alonso:2017bff,Alonso:2017uky,1674-1137-42-3-033104,Bonilla:2017lsq,Bhatia:2017tgo,Ellis:2017nrp,CHEN2018420,Faisel:2017glo,PhysRevD.97.115003,Bian:2017xzg,PhysRevD.97.075035,King:2018fcg,Duan:2018akc,Allanach:2018lvl,Allanach:2018odd}. Some models also have several abelian groups~\cite{Crivellin:2016ejn} leading to multiple $Z^\prime$s. Some other models~\cite{Kamenik:2017tnu,Camargo-Molina:2018cwu} generate BSM contributions to $C_L$ and $C_R$ with loop-level penguin diagrams. 

In Ref.~\cite{Allanach:2018lvl}, we introduced the Third Family Hypercharge
Model (TFHM). This model is based on a spontaneously broken  anomaly-free flavour-dependent
$U(1)_F$ symmetry, namely gauged third family hypercharge, and has the following desirable properties:
\begin{itemize}
  \item A $Z^\prime$ particle of several TeV in mass is predicted which can
    explain the NCBAs. The couplings $g_{sb}$ and $g_{\mu_L}$ are generated from the rotation between the weak and mass      eigenstates.
    \item The $Z^\prime$ does not appreciably couple to first or second family
      quarks (except to the second family quarks through the coupling $g_{sb}$), which 
    is hinted at by a number of experimental data; firstly, the absence of any
    similar neutral current anomalies in the semi-leptonic decays of lighter
    mesons such as kaons, pions, or charm-mesons; secondly, the absence of
    significant deviations
    with respect to the SM predictions for neutral meson mixing in the kaon
    and $B_d$ systems; and thirdly, the current absence of direct $Z^\prime$
    production in $pp$ collisions at the Large Hadron Collider (LHC), since the production cross-section would be enhanced by sizeable couplings of the $Z^\prime$ to valence quarks.
  \item The (3,3) entries of the up quark, down quark, and charged lepton Yukawa
  matrices were the only ones predicted to be non-zero at the renormalisable
  level. Small corrections to this picture are expected from non-renormalisable
  operators, but the model explains the hierarchical heaviness of the top and bottom
  quarks and the tau lepton. It also implies that the two CKM mixing angles involving the third family must be
  small, agreeing with current experimental measurements.
  \item The model is free of any gauge anomalies (including mixed or gravitational anomalies) without
  needing to introduce any additional chiral fermions beyond those of the SM
  (although sterile right-handed neutrinos may be added in order to provide a mechanism
  for neutrino mass generation).
  
  \end{itemize}
The charge assignment of the TFHM is shown in Table~\ref{tab:chargesTFHM}. The most up-to-date experimental bounds on the parameter space of the TFHM are presented in Ref.~\cite{Davighi:2019jwf}. 

\begin{table}
\begin{center}
\begin{tabular}{|ccc|}\hline
$F_{Q_i'}=0$ &  $F_{{u_R}_i'}=0$ & $F_{{d_R}_i'}=0$ \\ 
$F_{Q_3'}=1/6$ &  $F_{{u_R'}_3}=2/3$ & $F_{{d_R'}_3}=-1/3$ \\
$F_{L_i'}=0$ & $F_{{e_R}_i'}=0$ & $F_H=-1/2$ \\
$F_{L_3'}=-1/2$ & $F_{{e_R'}_3}=-1$ & $F_\theta$ \\ 
\hline
\end{tabular}
\caption{\label{tab:chargesTFHM} $U(1)_F$ charges of the fields in the original Third
  Family Hypercharge Model (TFHM), where $i \in \{1, 2\}$. All gauge
anomalies, mixed gauge anomalies and mixed gauge-gravity anomalies cancel. Under the SM gauge symmetry $SU(3)\times SU(2)_L
\times 
U(1)_Y$, the fields transform as $H \sim(1, 2, -1/2)$,
${Q_i}_L'\sim(3, 2, 1/6)$,
${L_i}_L'\sim(1, 2, -1/2)$, 
${u_i}_R'\sim(3, 1, 2/3)$,
${d_i}_R'\sim(3, 1, -1/3)$,
${e_i}_R'\sim(1, 1, -1)$, $\theta \sim (1,1,0)$. $F_\theta$ is left undetermined.}
\end{center}
\end{table}

\subsection{Motivation for extending the TFHM}

Despite these virtues, there is a somewhat ugly feature arising in the charged lepton sector of the model, as follows.
In order to transfer the $Z^\prime$ coupling from $\tau_L'$ to $\mu_L$ so that
the NCBAs may be fit, the
TFHM requires large mixing between the weak and mass eigenstates
of these two fields~\cite{Allanach:2018lvl}.
However, individual lepton numbers,
which are accidental symmetries of the SM, appear to be symmetries in Nature
to a good approximation, since experiments place strong upper bounds on lepton flavour
violating processes (e.g.\ in $\tau \rightarrow 3 \mu$ or $\mu \rightarrow e
\gamma$). Thus, introducing a flavour-changing interaction through large
charged lepton mixing is potentially dangerous from the point of view of these
bounds. 
Indeed, experimental constraints on $BR(\tau\rightarrow 3 \mu)$~\cite{PhysRevD.98.030001} 
place a tight bound on the coupling $g_{\mu\tau}\overline{\mu_L}\slashed{Z}^\prime \tau_L + H.c$.
This favours a mixing angle which is very close to $\pi/2$ between the second and third
family left-handed charged leptons.  
Such a mixing angle implies the renormalisable (3,3) Yukawa coupling for charged leptons must in fact be highly suppressed
with respect to the (2,3) and (3,2) Yukawa couplings (which, recall, can only arise from non-renormalisable operators given the charge assignment in the TFHM).
The TFHM model as presented in Refs.~\cite{Allanach:2018lvl,Davighi:2019jwf} has no explanation for this {\em per
  se}\/ because the (3,3) charged lepton Yukawa coupling should be present at
the renormalisable level and must therefore
be set to be small without explanation. From the outset, the model appears less natural because
of this; a deformed model which
does not require large 
$\mu_L-\tau_L$ mixing in order to obtain $g_{\mu_L} \neq 0$ or $g_{\mu_R} \neq 0$
would be more natural. 
In this paper, we will construct such an anomaly-free deformation of the TFHM, in
which the third family quarks and leptons {\em and}\/ the second family leptons are charged under
$U(1)_F$, which we shall see remedies the aforementioned ugly feature, while preserving
the successes of the TFHM\@. 

There is a second, albeit less troublesome, niggle in the TFHM setup.
If we were to assume that CKM mixing came from down quarks only,
the TFHM would obtain the wrong sign for $C_L \propto g_{sb} g_{\mu_L}$. Thus, additional CKM mixing
(of the opposite sign and roughly double the magnitude) must be invoked in the TFHM between
$t_L$ and $c_L$, allowing $g_{sb}$ to be of the correct sign {\em and}\/
magnitude. This is another feature that will be remedied in our deformation of the
TFHM, which will rather be compatible with purely down-quark CKM mixing (as
well as the case where there is also 
a contribution from the up quarks).

The resulting model, which we call the Deformed Third Family Hypercharge Model (DTFHM), is 
in these ways a more natural explanation of the NCBAs than the TFHM\@. Interestingly, we shall see that
the charge assignment in the DTFHM predicts contributions to both Wilson
coefficients  $C_L$ and $C_R$ (rather than just $C_L$, as predicted in the
original TFHM example case), in the particular combination
$C_L+\frac{4}{5}C_R$ (at least, in the most natural example case of the
DTFHM). To our knowledge, no model has been suggested to explain the NCBAs
with this particular ratio of Wilson coefficients. We find that such a
combination of operators can indeed fit the NCBA data much
  better than the SM.

In \S~\ref{sec:ttfhm}, we shall construct the TFHM deformation, calculating
the $Z^\prime$ couplings and the $Z-Z^\prime$ mixing therein. Then, in
\S~\ref{sec:pheno}, we examine the phenomenology of an {\em example case}\/ of the
model ({\em i.e.}\ with various simplifying assumptions about fermion
mixing). Firstly, the parameter space where the model fits the NCBAs is
estimated. Then other phenomenological bounds are examined, notably from $B_s$ mixing and the
measured lepton flavour universality of $Z$ couplings.
Direct $Z^\prime$ search constraints are calculated next, and we find that the
model has parameter space which evades all bounds but which explains the NCBAs
successfully. We summarise in \S~\ref{sec:sum}. In
  Appendix~\ref{sec:app}, we begin to sketch how some features of the example
  case might be predicted in a more complete and detailed model.

\section{The Deformed Third Family Hypercharge Model\label{sec:ttfhm}}

We deform the TFHM by allowing $U(1)_F$ charges (in the weak eigenbasis)
  not only for the third family of SM fermions, but also for the second family
  leptons, thus coupling the $Z^\prime$ directly to muons so that charged
  lepton mixing, and the lepton flavour violation (LFV) that it induces, can be
  small. In the spirit of bottom-up model building we shall not invoke any
  additional fields beyond those of the SM, the $Z^\prime$, and a flavon field
  whose r\^{o}le is to spontaneously break $U(1)_F$ at the scale of a few TeV by
  acquiring a non-zero vacuum expectation value (VEV).

To constrain our $U(1)_F$ charges we shall, as in the TFHM, require
  anomaly cancellation. This avoids the complication of including appropriate
  Wess-Zumino (WZ) terms to cancel anomalies in an otherwise anomalous
  low-energy effective field theory (EFT). Moreover, even if a specific set of
  anomalies could be cancelled at high energies by new UV physics, such as a set
  of heavy chiral fermions (from which the WZ terms must emerge as low-energy
  remnants), it would be difficult to give these chiral fermions large enough
  masses in a consistent framework, without prematurely breaking $SU(2)_L$.
  Thus, we require that our charge
  assignment is anomaly-free.  

\subsection{Anomaly-free deformation}

It turns out that the constraint of anomaly cancellation is strong enough to
uniquely determine the charge assignment in our deformation of the TFHM up to
a constant of proportionality. 
To derive this, we shall use the machinery
developed in Ref.~\cite{Allanach:2018vjg}, albeit in an especially simple
incarnation. We shall denote the $U(1)_F$ charge of field $M$ under $U(1)_F$
by $F_M$, where $M \in \{ Q_i,L_i,e_i,u_i,d_i,H \}$ and the index $i \in
\{1,2,3\}$ labels the family. In the DTFHM, non-zero charges are allowed only
for $M \in \{ Q_3,u_3,d_3,L_2,L_3,e_2,e_3,H \}$. We shall for now normalise
the gauge coupling $g_F$ such that all the $U(1)_F$ charges $F_M$, which are
necessarily rational numbers\footnote{We disallow the ratio of any two charges
being irrational since the charge assignment would then not fit into some
non-abelian unified group, which we expect will under-pin our model in the ultra-violet.}, are taken to be integers. 

There are six anomaly cancellation conditions (ACCs) which must hold, which guarantee the vanishing of all possible one-loop triangle diagrams involving at least one external $U(1)_F$ gauge boson, and two other external gauge bosons.
Four of these equations are linear in the charges, and together these
equations fix the third family quark charges to be proportional to their
hypercharges as in the TFHM\@. This, along with a choice for the constant of
proportionality, results in the charge assignments
\begin{equation}
F_{Q_3} = 1,\quad  F_{u_3} = 4,\quad F_{d_3} = -2,
\end{equation}
and also enforce
\begin{equation} \label{lepton plus}
F_{L_2} + F_{L_3} = -3, \quad F_{e_2} + F_{e_3} = -6.
\end{equation}
The remaining two ACCs are non-linear, one being quadratic and the other cubic. Following \cite{Allanach:2018vjg}, we recast these two equations in terms of the variables $F_{L-} = F_{L_2} - F_{L_3}$ and $F_{e-} = F_{e_2} - F_{e_3}$. We find that with this prudent choice of variables, the cubic ACC necessarily vanishes, and the quadratic one becomes simply
\begin{equation} \label{quad1}
F_{e-}^2 - F_{L-}^2 = 27.
\end{equation}
This equation is guaranteed to have at least one integer solution, because any
odd number $2m+1$ can be written as the difference of two consecutive squares, since $2m+1=(m+1)^2-m^2$. Thus, we have the solution.
\begin{equation}
14^2 - 13^2 = 27.
\end{equation}
Eq. (\ref{quad1}) has another integer solution, namely $6^2 - 3^2 =
27$.
However, setting $F_{e-} = 6$ and $F_{L-} = 3$, together with
  Eq. (\ref{lepton plus}), implies $F_{L_2} = F_{e_2} = 0$, and so this ``trivial
  branch" of solution just returns us to the TFHM charge assignment. It is
  straightforward to check that there are no other solutions to
  Eq. (\ref{quad1}) in which $F_{e-}$ and $F_{L-}$ are both integers.
Thus, choosing $F_{e-} = 14$ and $F_{L-} = 13$, we deduce the lepton charges:
\begin{equation}\label{U1charges}
F_{L_2} = +5, \quad F_{L_3} = -8, \quad F_{e_2} = +4, \quad F_{e_3} = -10.
\end{equation}
Hence we see that, given our assumptions, enforcing anomaly cancellation does
indeed fix a unique charge assignment.\footnote{Note that the charge
  assignment in Eq. (\ref{U1charges}) is only unique up to permutations of the
  family indices within each species; there are four such permutations, each
  corresponding to a different deformation of the TFHM\@. We choose the
  particular permutation in Eq. (\ref{U1charges}) for simple phenomenological
  reasons. Firstly, we choose $(F_{L_2},F_{L_3}) = (+5,-8)$ so that $F_{L_2}$
  and $F_{Q_3}$ have the same sign, allowing for the quark mixing to come from
  the down quarks only. Secondly, we choose the permutation
  $(F_{e_2},F_{e_3}) = (+4,-10)$ so that $|F_{L_2}|>|F_{e_2}|$, since fits to
  the NCBAs prefer a dominant coupling to left-handed muons, rather than
  right-handed muons. Indeed, if we were to choose the other
    permutation, {\em i.e.} $(F_{e_2},F_{e_3}) = (-10,+4)$, then the resulting
    combination of Wilson coefficients, $C_L-2C_R$, offers a fit to the NCBA
    data of a bad quality, similar to the SM - see the green dashed line in
    the left-hand plot of Fig.~\ref{fig:dcsq}.
}

For the rest of the paper, we divide all the $F_M$ charges by $6$, so that the
quarks and Higgs doublet
have their charges equal to the usual 
hypercharge assignment. The $U(1)_F$ charge
assignment of the DTFHM, in the weak eigenbasis, is then listed in Table~\ref{tab:chargesDTFHM}.  
\begin{table}
\begin{center}
\begin{tabular}{|ccc|}\hline
$F_{Q_1'}=0$ &  $F_{{u_R}_1'}=0$ & $F_{{d_R}_1'}=0$ \\
$F_{Q_2'}=0$ &  $F_{{u_R}_2'}=0$ & $F_{{d_R}_2'}=0$ \\
$F_{Q_3'}=1/6$ &  $F_{{u_R'}_3}=2/3$ & $F_{{d_R'}_3}=-1/3$ \\
$F_{L_1'}=0$ & $F_{{e_R}_1'}=0$ & $F_H=-1/2$  \\
$F_{L_2'}=5/6$ & $F_{{e_R}_2'}=2/3$ & $F_\theta$\\
$F_{L_3'}=-4/3$ & $F_{{e_R'}_3}=-5/3$ & 
  \\ \hline
\end{tabular}
\caption{\label{tab:chargesDTFHM} $U(1)_F$ charges of the fields in the Deformed Third
  Family Hypercharge Model (DTFHM). All gauge
anomalies, mixed gauge anomalies and mixed gauge-gravity anomalies cancel with
this charge assignment, which has been previously listed in an `Anomaly-Free
Atlas' in Ref.~\protect\cite{Allanach:2018vjg,b_c_allanach_2018_1478085}. At this stage, $F_\theta$ is left undetermined.}
\end{center}
\end{table}

Before we proceed to flesh out the details of this model, let us make a few
comments. Firstly, $F_{L_2}$ (and $F_{e_2}$) now have the same sign as
$F_{Q_3}$. This means that we may assume the CKM mixing comes from the down
quarks only, which would produce a coupling $g_{sb}\propto V_{tb} V_{ts}^\ast
F_{Q_3'}$, and obtain $C_L \propto g_{sb}g_{\mu_L}<0$ (neglecting small imaginary parts in the CKM matrix elements), 
the correct sign for fitting
the NCBAs. Secondly, the magnitude of the lepton charges are large compared
with $F_{Q_3}$, which shall make the constraints from $B_s-\overline{B_s}$
mixing easier to satisfy while simultaneously providing a good fit to the
NCBAs. Thirdly, as mentioned above, the $Z^\prime$ coupling to the muon is no
longer left-handed, but is now proportional to $C_L +
\frac{4}{5}C_R$.


Finally, let us discuss the implications of this new charge assignment for the Yukawa sector of the model.
With the charge assignment in Table~\ref{tab:chargesDTFHM}, the only renormalisable Yukawa couplings are 
\begin{equation}
{\mathcal L}=-Y_t  {\overline{{Q_3}_L'}} H t_R' - Y_b \overline{{Q_3'}_L} H^c b_R'+ H.c, \label{DTFHM yuks}
\end{equation}
where we suppress gauge indices
and $H^c=(H^+,\ -{H^0}^\ast)^T$.
In contrast to the TFHM, all Yukawa couplings for the charged leptons are now
banned at the renormalisable level, even the (3,3) element. So there is no
expectation for a heavy tauon in this theory, whose mass would therefore, like
the first and second family fermions, arise from non-renormalisable
operators. We find this palatable given $m_\tau \simeq 1.7\ \mathrm{GeV} \ll
m_t$. Indeed, $m_\tau$ is closer to the charm mass, $m_c \simeq 1.3\
\mathrm{GeV}$ (which like other second family fermion masses must also arise at the non-renormalisable level) than it is to either of the third family quark masses.

In this model, one would still expect the bottom and top quarks to be
hierarchically heavier than the lighter quarks, and expect small CKM angles
mixing the first two families with the third. One would not necessarily expect
the CKM mixing between the first two families to be small (as indeed it is
not), given the approximate $U(2)$ symmetry in the light quarks, as is also
the case in the TFHM and many other models. We require a small renormalisable
parameter $Y_b \sim 1/40$ in order to fit the bottom mass. 

\subsection{Neutrino masses}

If we augment the SM fermion content by three right-handed sterile neutrinos
$\nu^\prime_{iR}$, $i\in\{1,2,3\}$, which are uncharged under all of SM$\times
U(1)_F$, then, given the charge assignment in Table~\ref{tab:chargesDTFHM},
one cannot write down any renormalisable Yukawa couplings for
neutrinos. Nonetheless, just as for the charged leptons and the first and
second family quarks, we expect effective Yukawa operators for the neutrinos,
of the form $\overline{{L_i'}_L} H \nu'_{jR}$, to arise from higher-dimension
operators, for example involving insertions of the flavon field. Moreover, once
we include three right-handed sterile neutrinos, then we should also include a
generic 3 by 3 matrix of Majorana mass terms in the Lagrangian, which
correspond to super-renormalisable dimension-3 operators of the form
$\overline{{\nu_i'}_R^c} \nu'_{jR}$, whose dimensionful mass parameters reside
at some {\em a priori}\/ decoupled heavy mass scale. Thus, employing the same
coarse reasoning with which we discussed quark and lepton masses, it is
natural to expect a spectrum of three light neutrinos within our model,
together with three very heavy right-handed counterparts, arising from a
see-saw mechanism. 

In the remainder of this Section, we complete our description of this model by discussing first the neutral gauge boson mass mixing, which results from the Higgs being charged under both the electroweak symmetry and under $U(1)_F$, and second the coupling of the $Z^\prime$ to the fermion sector. These aspects are similar to the setup of the original TFHM, as described in Ref.~\cite{Allanach:2018lvl}.

\subsection{Masses of gauge bosons and $Z-Z^\prime$ mixing} \label{ZZpmixing}

The mass terms for the neutral gauge bosons are of the form
$\mathcal{L}_{N,\text{mass}} = \frac{1}{2}{\bm A'_{\mu}}^T
\mathcal{M}^2_N {\bm A'_{\mu}}$, where ${\bm
  A'_{\bm \mu}}=(B_\mu,W_\mu^3,X_\mu)^T$,\footnote{Here, the prime on ${\bm A'_\mu}$
  denotes that the gauge fields are in the $SU(3) \times SU(2)_L \times U(1)_Y
  \times U(1)_F$ eigenbasis.}  and
\begin{equation}
\mathcal{M}^2_N=\frac{v_F^2}{4}\left( \begin{array}{ccc}
r^2 {g'}^2 & -r^2 gg' & r^2 g'g_F \\
-r^2 gg' & r^2g^2 & -r^2gg_F \\
r^2 g'g_F & -r^2gg_F & g_F^2(4 F_\theta^2+r^2) \\
\end{array}\right),
\end{equation}
where $r\equiv v/v_F\ll 1$ is the ratio of the VEVs, and $F_\theta$ is the
$U(1)_F$ charge of the flavon $\theta$. The mass basis of physical neutral
gauge bosons is defined via $(A_\mu,Z_\mu,Z^{\prime}_\mu)^T\equiv{\bm
  A}_{\bm \mu}=O^T {\bm A'_{\mu}}$, where 
\begin{equation}
O=
\left( \begin{array}{ccc}
\cos\theta_w & -\sin\theta_w \cos\alpha_z & \sin\theta_w \sin\alpha_z \\
\sin\theta_w & \cos\theta_w \cos\alpha_z & -\cos\theta_w \sin\alpha_z \\
0 & \sin\alpha_z & \cos\alpha_z \\
\end{array}\right), \label{orthogonal}
\end{equation}
where $\theta_w$ is the Weinberg angle (such that $\tan\theta_w=g'/g$). In the (consistent) limit that $M_Z/M_Z^\prime\ll 1$ and $\sin\alpha_z \ll 1$, the masses of the heavy neutral gauge bosons are given by
\begin{equation}
M_Z\approx \frac{M_W}{\cos\theta_w}=v\frac{\sqrt{g^2+{g'}^2}}{2}, \qquad
M_{Z^\prime}\approx g_F v_F F_\theta, 
\label{gbMasses}
\end{equation}
where $M_W =gv/2 $, and the $Z-Z^\prime$ mixing angle is
\begin{equation}
\sin\alpha_z \approx \frac{g_F}{\sqrt{g^2+{g'}^2}}\left(\frac{M_Z}{M_Z^\prime}\right)^2. \label{mixing}
\end{equation}
Recall that we expect
$v_F \gg v$, so that the $Z^\prime$ is indeed expected to be much heavier than the electroweak gauge bosons of the SM.

\subsection{$Z^\prime$ couplings to fermions \label{zPcouplings}}

We begin with the couplings of the $U(1)_F$ gauge boson $X_\mu$ to fermions in
the Lagrangian in the weak (primed) eigenbasis
\begin{eqnarray}
{\mathcal L}_{X \psi} &=& -
g_F \left( 
\frac{1}{6}{\overline{{Q_3'}_L}} \slashed{X} {Q_3'}_L +\frac{2}{3}
{\overline{{u_3'}_R}} \slashed{X} {u_3'}_R -\frac{1}{3}
{\overline{{d_3'}_R}} \slashed{X} {d_3'}_R \right. \nonumber
\\ && \left.
+\frac{5}{6}
{\overline{{L_2'}_L}} \slashed{X} {L_2'}_L +\frac{2}{3}
{\overline{{e_2'}_R}} \slashed{X} {e_2'}_R  \right. \nonumber
\\ && \left.
-\frac{4}{3}{\overline{{L_3'}_L}} \slashed{X} {L_3'}_L -\frac{5}{3}
{\overline{{e_3'}_R}} \slashed{X} {e_3'}_R 
\right), \label{Zpcouplings}
\end{eqnarray}
where
$g_F$ is the $U(1)_F$ gauge coupling.
Writing the weak eigenbasis fields as 
3-dimensional vectors in family space ${\bf u_R}'$, ${\bf Q_L}'=({\bf
  u_L}',\ {\bf d_L}')$, ${\bf e_R}'$,
${\bf d_R}'$, ${\bf L_L}'=({\bm{\nu}_L}', {\bf e_L}')$, we define the 3 by 3 unitary matrices $V_P$, where
$P \in \{u_R,\ d_L,\ u_L,\ e_R,\ u_R,\ d_R,\ \nu_L,\ e_L \}$ to transform between the weak
eigenbasis and the mass (unprimed) eigenbasis\footnote{The transposes on the
  vectors (e.g.\ ${\bf P}^T$) denote that the result is to be thought of as a
  {\em column}\/ vector in family space.}:
\begin{equation}
  {{\bf P}'}^T = V_P {\bf P}^T.
  \end{equation}
The CKM matrix is $V=V_{u_L}^\dag V_{d_L}$ and the Pontecorvo-Maki-Nakagawa-Sakata (PMNS) matrix is
$U=V_{\nu_L}^\dag V_{e_L}$. Re-writing Eq.~\ref{Zpcouplings} in the mass
eigenbasis,
\begin{eqnarray}
{\mathcal L}_{Z^\prime \psi} &=& -
g_F \left( 
\frac{1}{6}{\overline{{\bf d_L}}} \Lambda^{(d_L)}_\xi \slashed{Z}^\prime {\bf d_L} +
\frac{1}{6}{\overline{{\bf u_L}}} \Lambda^{(u_L)}_\xi \slashed{Z}^\prime {\bf u_L} 
\right. \nonumber \\ && \left.
+\frac{2}{3}{\overline{{\bf u_R}}} \Lambda^{(u_R)}_\xi \slashed{Z}^\prime {\bf u_R}
-\frac{1}{3} {\overline{{\bf d_R}}} \Lambda^{(d_R)}_\xi \slashed{Z}^\prime {\bf d_R}
\right. \nonumber \\ && \left.
+\frac{5}{6} {\overline{{\bf e_L}}} \Lambda^{(e_L)}_\Omega \slashed{Z}^\prime {\bf e_L} 
+\frac{5}{6} {\overline{{\bm{\nu}_L}}} \Lambda^{(\nu_L)}_\Omega \slashed{Z}^\prime {\bm{\nu}_L}
\right. \nonumber \\ && \left.
+\frac{2}{3} {\overline{{\bf e_R}}} \Lambda^{(e_R)}_\Psi \slashed{Z}^\prime {\bf e_R}
\right), \label{Zpcoupmass}
\end{eqnarray}
up to small terms $\sim {\mathcal O}(M_Z^2/M_{Z^\prime}^2)$ induced by the
$Z-Z^\prime$ mixing. 
We have defined the 3 by 3 matrices in family space $\Lambda^{(P)}_\zeta = V_P^\dag \zeta V_P$, where $\zeta \in 
\{ \xi, \Omega, \Psi\}$ and
  \begin{equation}
    \xi= \left( \begin{array}{ccc}
      0 & 0 & 0 \\
      0 & 0 & 0 \\
      0 & 0 & 1 \\
    \end{array}\right), \qquad
    \Omega= \left( \begin{array}{ccc}
      0 & 0 & 0 \\
      0 & 1 & 0 \\
      0 & 0 & -\frac{8}{5} \\
    \end{array}\right), \qquad    
    \Psi= \left( \begin{array}{ccc}
      0 & 0 & 0 \\
      0 & 1 & 0 \\
      0 & 0 & -\frac{5}{2} \\
    \end{array}\right). 
    \end{equation}
In order to make further progress with phenomenological analysis, we must fix
the fermion mixing matrices $V_P$. 
The $Z^\prime$ boson couples to both left-handed and right-handed muons, as can be seen by reference to
Eq.~\ref{Zpcoupmass} 
and the non-zero (2,2)
entries of $\Psi$ and $\Omega$. However, in order to fit the NCBAs, we require
a coupling of the $Z^\prime$ 
to $\overline{s_L} b_L$. This implies that $(V_{d_L})_{23} \neq 0$.
The $Z^\prime$ will then, once integrated out of the effective field theory,
induce a $(\overline{s} b) (\bar \mu \mu)$ effective operator which, we shall
show below, can explain the NCBAs.
The $Z^\prime$ also mediates other flavour-changing neutral
currents, and so will be subject to various bounds upon its flavour-changing or flavour
non-universal couplings. This will translate to bounds upon the various
entries of the $V_P$.

\subsection{Example case \label{sec:eg}}
We shall here construct an example of the set of $V_P$ that is not obviously
ruled out {\em a priori}, for further phenomenological analysis.
We
shall assume that the currently measured CKM quark mixing is due to the
down quarks, thus $V_{d_L}=V$, $V_{u_L}=1$.
Explicitly, this yields the following matrix of couplings to down quarks
\begin{equation} \label{down couplings}
\Lambda^{(d_L)}_{\xi} = \left( \begin{array}{ccc}
|V_{td}|^2 & V_{td}^* V_{ts} & V_{td}^* V_{tb} \\
V_{td} V_{ts}^* & |V_{ts}|^2 & V_{ts}^* V_{tb} \\
V_{tb}^* V_{td} & V_{tb}^* V_{ts} & |V_{tb}|^2 \\
\end{array} \right).
\end{equation}
We shall also assume that the
observed PMNS mixing is due solely to the neutrinos,
{\em i.e.}\ $V_{\nu_L}=U^\dagger$, $V_{e_L}=1$. We note that (in contrast to the original 
TFHM), despite there being no charged lepton mixing, there is a $Z^\prime$
coupling to muons.
For simplicity and definiteness, we choose $V_{u_R}=1=V_{d_R}=V_{e_R}$.
The assumed alignment of the charge lepton weak basis with the mass basis
ensures no lepton flavour violation (LFV), which is very tightly constrained
by experimental measurements (in particular $\tau \rightarrow 3\mu$, and $\mu
\rightarrow e\gamma$).
The example case corresponds to some strong (but reasonable) assumptions about the $V_P$, which
may not hold in reality. In the future, we may perturb away from this particular
example case, but it will suffice for a first study of viable parameter
space and relevant  direct $Z^\prime$ search limits from the LHC\@.
In what follows, we shall refer to this example case of the DTFHM as the
`DTFHMeg'.

The assumptions on the mixing matrices $\{V_P\}$ are rather strong, and one may ask what would be
  required of some more detailed model in order to obtain them. For example, to obtain $V_{u_L}=V_{u_R}=1$, 
  we require that the predicted form of the up-quark Yukawa
  matrix be diagonal in the weak eigenbasis. In Appendix~\ref{sec:app}, we
  sketch how this might be achieved in a more detailed
  Froggatt-Nielsen model.


\section{Phenomenology of the Example Case \label{sec:pheno}} 

In this section, we will go through the most relevant phenomenological limits on the DTFHMeg in turn,
concluding with a discussion of the combination of them all. The phenomenology that we discuss
is only sensitive to $F_\theta$ and $v_F$ through $M_{Z^\prime}$ in
Eq.~\ref{gbMasses}. We shall leave each undetermined and use $M_{Z^\prime}$ as
the independent variable instead.

\subsection{NCBAs}

\begin{figure*}[ht]
\begin{center}
\unitlength=\textwidth
\includegraphics[width=0.35 \textwidth]{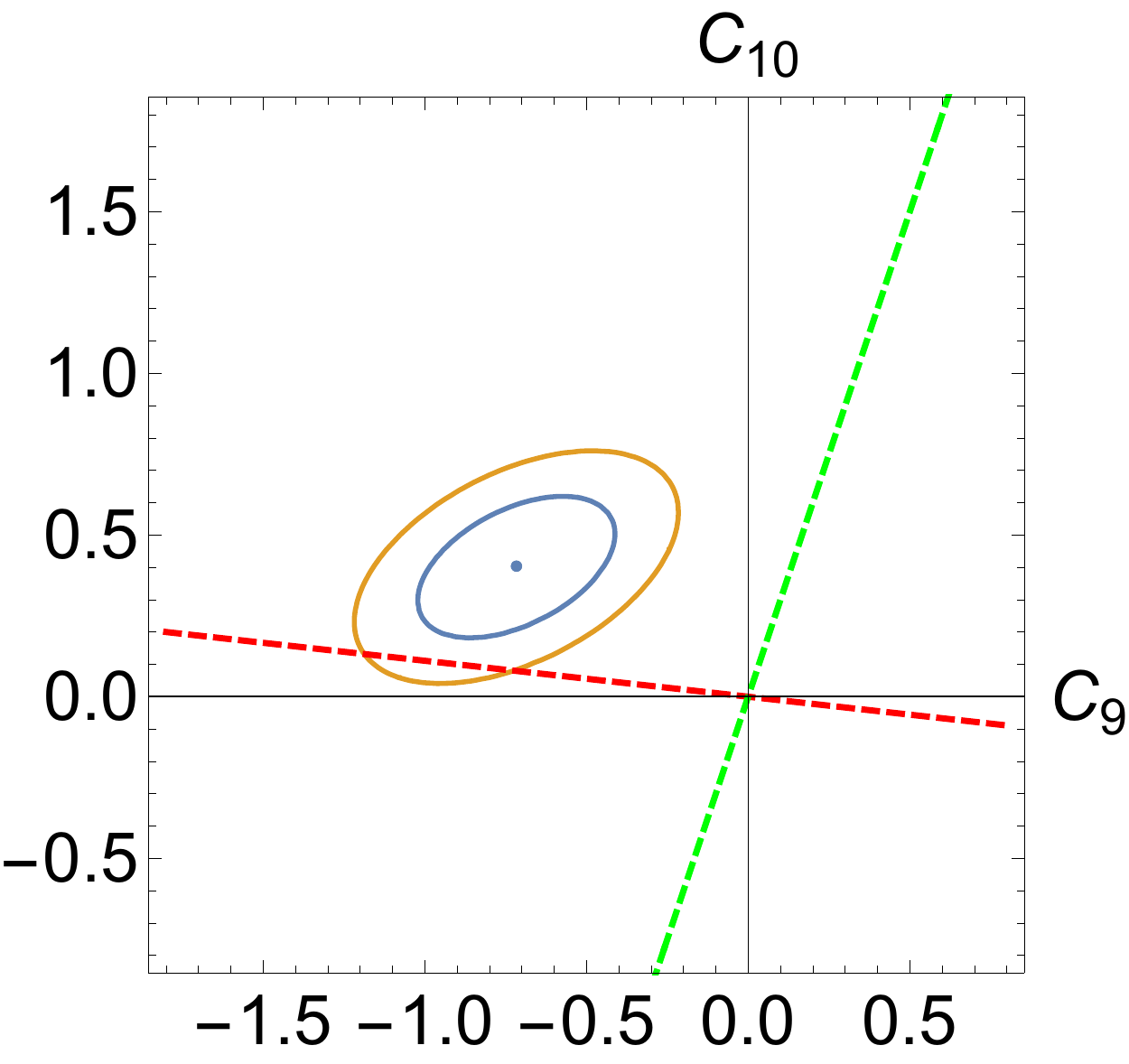}
\includegraphics[width=0.45 \textwidth]{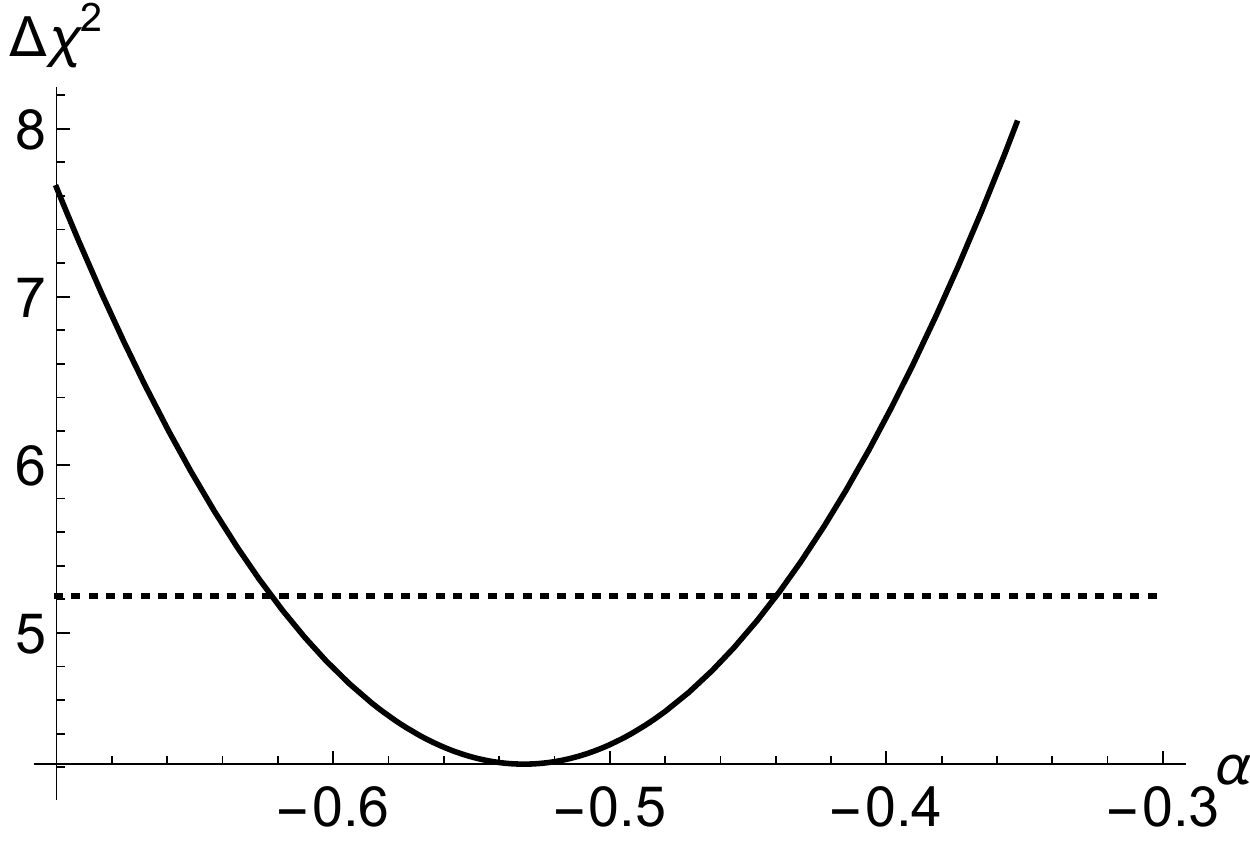}
\end{center}
\caption{Our digitisation of the fits of
Ref.~\cite{Aebischer:2019mlg}. {\em Left}\/: The point shows the best-fit in $(C_9,
C_{10})$ space, surrounded by
68$\%$ (inner) and 95$\%$ (outer) $CL$ regions. The red dashed line shows the trajectory of
our model, which predicts that $C_9 = -9 C_{10}$. We see that
  this ratio of Wilson coefficients is capable of fitting the NCBA data at the
  1.5$\sigma$ level. For reference, we also include (green
  dashed line) the trajectory of models corresponding to the same anomaly-free
  charge assignment, but with the lepton families permuted such that
  $F_{L_2\prime}$ and $F_{e_2\prime}$ are in the ratio $-2:1$ or $1:-2$. Such
  a model only offers a bad quality fit to the NCBAs, similar to the
  SM. {\em Right}\/: Shows 
$\Delta \chi^2(\alpha)$ as a function of $\alpha$ along the
red dashed line ({\em i.e.}\ $C_9=-9 C_{10}$).
The horizontal dotted line shows $\Delta \chi^2$ of unity above the 
best-fit value, and is used to calculate the 1$\sigma$ uncertainties on $\alpha$. \label{fig:dcsq}} 
\end{figure*}

From the global fit to $C_9$ and $C_{10}$ in Ref.~\cite{Aebischer:2019mlg}
(the left-hand panel of Fig.~1), we 
extract the fitted BSM contributions from the 68$\%$ CL
ellipse 
\begin{equation}
  \left( \begin{array}{c} C_9 \\C_{10} \end{array}\right) =
       {\bf c} +        
  \frac{s_1}{\sqrt{2.3}} {\bf v_1}+ \frac{s_2}{\sqrt{2.3}} {\bf v_2},
\end{equation}
where\footnote{The $1/\sqrt{2.3}$ factors come from the fact that the combined
fit is in 2 dimensions, so Ref.~\cite{Aebischer:2019mlg} plots the 68$\%$
confidence level region as $\Delta \chi^2=2.3$ from the best-fit point.}
${\bf c}=(-0.72,\ 0.40)^T$, 
${\bf v_1}=(0.29,\ 0.15)^T$, ${\bf v_2}=(-0.08,\ 0.16)^T$ is
orthogonal to ${\bf   v_1}$ and $s_1$, $s_2$ are independent
one-dimensional Gaussian probability 
density functions with mean zero and unit standard deviation.
We are thus working in the approximation that the fit yields a 2-dimensional
Gaussian PDF near the likelihood maximum.
We plot our characterisation of the 68$\%$ and 95$\%$ error ellipses in
Fig.~\ref{fig:dcsq} (left). Overlaying it on top of Fig.~1 of
Ref.~\cite{Aebischer:2019mlg} shows that this is a good approximation in the
vicinity of the best-fit point. 

The best-fit point has a $\chi^2$ of some 42.2 units {\em less}\/ than the SM~\cite{Aebischer:2019mlg}.  
We have $C_9=C_L + C_R$ and $C_{10}=C_R - C_L$, so, for the DTFHMeg in which $C_L=\alpha$ and $C_R
= 4/5 \alpha$, we have $(C_9,\ C_{10})={\bf d}(\alpha) \equiv \alpha(9 /5,\ -1/5)$.
We may use the orthogonality of ${\bf v_1}$ and ${\bf v_2}$ to solve for
\begin{equation}
  s_i(\alpha) = \frac{\sqrt{2.3}}{|{\bf v_i}|^2}{\bf v_i}\cdot \left( {\bf d}(\alpha)
    - {\bf c}\right),
\end{equation}
where $i \in \{1,2\}$.
The
value of $\Delta \chi^2$ that we extract from the fit is then the difference
in $\chi^2$ between our fit and the best fit point in $(C_9,\ C_{10})$ space:
\begin{equation}
\Delta \chi^2(\alpha) = s_1^2(\alpha)+s_2^2(\alpha). \label{dc2}
  \end{equation}
The value of $\alpha$ which minimises this function
($\alpha_{\text{min}}$) 
is the best-fit value and the places where it 
crosses $\Delta \chi^2(\alpha_{\text{min}})+1$ 
yield the $\pm 1\sigma$ estimate 
for its uncertainty under the hypothesis that the model is correct, {\em i.e.} 
\begin{equation}
  \alpha = -0.53 \pm 0.09. \label{alphaFit}
\end{equation}
The SM lies at $\alpha=0$ and so the fit is $5.9\sigma$
  away from it, of comparable quality to the two-parameter fit of $C_9$ and
  $C_{10}$ (which is 6.5$\sigma$ away from the SM point).
$\Delta \chi^2(\alpha)$ is plotted in the vicinity of the minimum in
Fig.~\ref{fig:dcsq} (right). 

This minimum is obtained at a higher $\Delta \chi^2(\alpha_{\text{min}})=4.2$ as compared
to the unconstrained fit to $(C_9,\ C_{10})$, for one parameter fewer, {\em i.e.}\ one additional degree of freedom. The model still
constitutes a good fit to the NCBAs, having a best-fit $\chi^2$ value 38.0
lower than the SM.

The couplings in the DTFHMeg relevant to a new physics contribution to
the $(\bar bs) (\bar \mu \mu)$ vertices are
\begin{equation}
  {\mathcal L}_{bs\mu\mu}= -g_F \left[ \frac{5}{12}\overline{\mu_L} \slashed{Z}^\prime
    \mu_L + \frac{1}{3} \overline{\mu_R} \slashed{Z}^\prime \mu_R +
    \frac{V_{ts}^\ast V_{tb}}{6} \overline{s_L}\slashed{Z}^\prime b_L +H.c.
  \right], \label{bsmm2}
\end{equation}
where `$+H.c.$' signifies that we are to add the Hermitian conjugate copies of
all terms in the square brackets.
 By reference
to Eqs.~\ref{NCBAZ} and~\ref{bsmm2}, we identify $g_{sb} = V_{ts}^\ast V_{tb} g_F/6$,
$g_{\mu_L}=5g_F/6$ and $g_{\mu_R}=2g_F/3$. 
Using $V_{ts}^\ast V_{tb} \approx -0.04$ and matching $C_L$ to $\alpha$'s fit
value in Eq.~\ref{alphaFit}, we obtain
\begin{equation}
  0.22 \leq g_F \frac{1 \text{~TeV}}{M_{Z^\prime}} \leq 0.31. \label{gm}
  \end{equation}
as the two sigma (95$\%$ CL) fit to the NCBAs.

\subsection{$Z^\prime$ width}
The partial width of a $Z^\prime$ decaying into a massless fermion $f_i$ and massless
anti-fermion $\bar f_j$ is 
$\Gamma_{ij}=C/(24 \pi) |g_{ij}|^2 M_{Z^\prime}$, where
$g_{ij}$ is the coupling of the $Z^\prime$ to $f_i \bar f_j$, and $C=3$ for
coloured fermions ($C=1$ otherwise). In the limit that $M_{Z^\prime}$ is much 
larger than the top mass, we may approximate all fermions as being massless. Summing over all fermion species, we obtain that the
total width $\Gamma$ satisfies $\Gamma/M_{Z^\prime}=5 g_F^2 / (12 \pi)$.
The model is non-perturbative when this quantity approaches unity, {\em i.e.}\ $g_F
\sim  \sqrt{12 \pi/5}=2.7$. Eq.~\ref{gm} implies that to avoid this
non-perturbative r\'{e}gime requires
$M_{Z^\prime} \lesssim 12.5$ TeV. 
The $Z^\prime$ in this model decays with the following branching ratios: 18$\%$ into quarks of
various flavours, 11$\%$ into muons, 46$\%$ into tauons, and 25$\%$ into neutrinos. 

We observe that these branching ratios are a significant departure from those
in the original TFHM; in particular, the branching ratios into quark pairs
(predominantly tops and bottoms) is much reduced, and the branching ratio into
neutrinos and tauons is much enhanced in the DTFHM\@. This is because of the
significantly larger lepton charges in this model (which, recall, were fixed
by anomaly cancellation), and the fact that the coupling to left-handed tauons
no longer needs to be transferred into a coupling to left-handed muons in the
example case.  

\subsection{Neutral meson mixing}

The most recent constraint coming from comparing $B_s$ mixing predictions from 
lattice data and sum rules~\cite{King:2019lal} with experimental measurements~\cite{Amhis:2016xyh} yields~\cite{Allanach:2019mfl}
$|g_{sb}| \leq
M_{Z^\prime}/(194\text{~TeV})$.
$B_s$ mixing thus usually places a strong constraint upon $Z^\prime$ models that
explain the NCBAs~\cite{DiLuzio:2017fdq}. Substituting for $g_{sb}$, we obtain
\begin{equation}
  g_F \frac{1 \text{~TeV}}{M_{Z^\prime}} < 0.77,
 \label{bsmix}
\end{equation}
which we see is satisfied by the whole $2\sigma$ range favoured by a fit to the NCBAs
in Eq.~\ref{gm}.

The flavour-changing couplings of the $Z^\prime$ to down quarks, given in Eq.~\ref{down couplings} in
our example case, also produce corrections beyond the Standard Model to the mixings of other neutral mesons, specifically to kaon and $B_d$ mixing. For the DTFHMeg, we compute the 95\% CL bound from neutral kaon mixing to be $g_F \left(1 \text{~TeV}/M_{Z^\prime}\right) < 1.46$, while that from $B_d$ mixing is $g_F \left(1 \text{~TeV}/M_{Z^\prime}\right) < 0.82$, where in both cases we have used the constraints presented in Ref.~\cite{Charles:2013aka}. Thus, the bound from $B_s$ mixing given above turns out to be the strongest of the three.

\subsection{$Z$ boson lepton flavour universality}\label{LFU}

Here, we follow Ref.~\cite{Allanach:2018lvl} to compute the bound coming from 
lepton flavour universality measurements of $Z$ boson couplings (with the difference that
here, we must also
include the contribution from $\mu_R$).
The LEP measurement is:
\begin{equation}
R_{\text{LEP}} =0.999\pm 0.003,
\qquad R\equiv\frac{\Gamma(Z\rightarrow e^+e^-)}{\Gamma(Z\rightarrow \mu^+\mu^-)}. \label{LEP}
\end{equation}
We compute the prediction for this in our model by evaluating the following ratio of partial widths,
\begin{equation}
R_{\text{model}} = \frac{|g_Z^{e_L e_L}|^2+ |g_Z^{e_R e_R}|^2}{|g_Z^{\mu_L \mu_L}|^2+|g_Z^{\mu_R \mu_R}|^2},
\end{equation}
where $g_Z^{ff}$ is the coupling of the physical $Z$ boson to the fermion
anti-fermion pair
$f\bar f$. One can obtain the couplings $g_Z^{ff}$ in the DTFHMeg by first writing down the terms
in the Lagrangian which couple the first and second family charged leptons to the neutral bosons $B$,
$W^3$, and $X$:
\begin{eqnarray}\mathcal{L}_{l Z^\prime}&=&-\overline{e_L}\left(-\frac{1}{2}g
\slashed{W}^3-\frac{1}{2}g'\slashed{B}\right) e_L -\overline{e_R}\left(-g'\slashed{B}\right) e_R \nonumber \\ &&
-\overline{\mu_L}\left(-\frac{1}{2}g \slashed{W}^3-\frac{1}{2}g'\slashed{B}+\frac{5}{6}g_F\slashed{X}\right) \mu_L \nonumber \\ &&
-\overline{\mu_R}\left(-g'\slashed{B}+\frac{2}{3}g_F\slashed{X}\right) \mu_R, \label{eq:charged lepton couplings}
\end{eqnarray}
and then inserting ${\bm A_\mu}'=O {\bm A_\mu}$ (where $O$ is given in
Eq.~\ref{orthogonal}) to rotate to the mass basis. 
To leading order in 
$\sin\alpha_z$, we find 
\begin{eqnarray}
\begin{aligned}
g_Z^{e_L e_L} &=-\frac{1}{2}g\cos\theta_w+\frac{1}{2}g'\sin\theta_w,\\
g_Z^{e_R e_R}&=g'\sin\theta_w,\\
g_Z^{\mu_L \mu_L} &=-\frac{1}{2}g\cos\theta_w+\frac{1}{2}g'\sin\theta_w+\frac{5}{6}g_F\sin\alpha_z,\\
g_Z^{\mu_R \mu_R} &= g'\sin\theta_w + \frac{2}{3}g_F\sin\alpha_z.
\end{aligned}
\end{eqnarray}
The SM prediction ({\em i.e.} $R=1$) is
recovered by taking $\alpha_z$ to zero. Within our model, we may expand $R_{\text{model}}$ to leading order in
$\sin\alpha_z$ to obtain
\begin{eqnarray}
R_{\text{model}} &=& 1
+\frac{2}{3}\frac{g_F(5g\cos\theta_w-13g'\sin\theta_w)\sin\alpha_z}{(g\cos\theta_w-g'\sin\theta_w)^2+4{g'}^2\sin^2\theta_w} \nonumber \\ 
&=& 1+2.6 g_F^2 \left(\frac{M_Z}{M_{Z^\prime}}\right)^2, 
\end{eqnarray}
having substituted in Eq.~\ref{mixing} for $\sin\alpha_z$, and the central experimental values $g=0.64$ and $g'=0.34$. 
Comparison with the upper 
LEP limit in Eq.~\ref{LEP}, at the $95\%$ CL,
yields the $Z$ boson lepton 
flavour universality constraint from LEP (which we henceforth refer to as the LEP LFU bound):
\begin{equation}
g_F^2 \left(\frac{M_Z}{M_{Z^\prime}}\right)^2< 0.0019 \Rightarrow
g_F \frac{1\text{~TeV}}{M_{Z^\prime}} < 0.48, \label{LEPLFU}
\end{equation}
which is satisfied by the entire range favoured by current fits to
NCBAs in Eq.~\ref{gm}.

One might have expected that, due to the enhanced $Z^\prime$ couplings to muons,
the LEP LFU bound would be more aggressive in the DTFHM than in the TFHM\@. 
However, in the DTFHM, a partial cancellation occurs between the contributions to $R_{\text{model}}$ coming from 
$g_Z^{\mu_L \mu_L}$ and $g_Z^{\mu_R \mu_R}$. 
This does not occur in the
original TFHM, in which the coupling of the $Z^\prime$ (and thus
of the $Z$, after $Z-Z^\prime$ mixing) to muons is purely left-handed. Due to
this partial cancellation, this constraint from LEP LFU in the DTFHMeg is
somewhat less aggressive than it would be otherwise, ending up very close to
that of the TFHM example case.

\subsection{Invisible Width of the $Z$ Boson}
$Z^\prime$ couplings contribute to the  invisible
  width of the $Z$ boson $\Gamma_{\text{inv}}$ beyond the SM via $Z-Z^\prime$ mixing and the
  $Z^\prime$ coupling to 
  neutrinos.
  Experimental constraints upon it are~\cite{ALEPH:2005ab}
  \begin{equation}
    {\Gamma}^{(\text{exp})}_{\text{inv}}=499.0 \pm 1.5 \text{~MeV} \label{gzexp},
  \end{equation}
whereas the SM prediction from the decay into $\overline{\nu_e} \nu_e$,
  $\overline{\nu_\mu} \nu_\mu$ and $\overline{\nu_\tau} \nu_\tau$
  is ${\Gamma}^{(\text{SM})}_{\text{inv}}=501.44$~MeV~\cite{PhysRevD.98.030001}.
  Thus, we may constrain any new physics contribution to be
  \begin{equation}
    \Delta \Gamma^{(\text{exp})} = {\Gamma}^{(\text{exp})}_{\text{inv}} - {\Gamma}^{(\text{SM})}_{\text{inv}} =
    -2.5 \pm 1.5 \text{~MeV}. \label{dgamexp}
  \end{equation}
  The terms in the Lagrangian that couple the $Z$ boson to neutrinos are, to leading
  order in $\sin \alpha_z \ll 1$,
  \begin{eqnarray}
    {\mathcal L}_{\bar \nu \nu Z} &=&-
     \frac{g}{2 \cos \theta_w} \overline{{\nu_L'}_e} \slashed{Z} P_L {\nu_L'}_e
    \nonumber \\ &&
    - \overline{{\nu_L'}_\mu}\left( \frac{g}{2 \cos \theta_w} + \frac{5}{6} g_F
    \sin \alpha_z \right)\slashed{Z} {\nu_L'}_\mu \nonumber \\ &&
    - \overline{{\nu_L'}_\tau}\left( \frac{g}{2 \cos \theta_w} - \frac{4}{3} g_F
    \sin \alpha_z \right)\slashed{Z} {\nu_L'}_\tau. \label{Znunu}
  \end{eqnarray}
In collider experiments the neutrinos are not observed and so their flavour is not measured.
We may therefore compute $\Delta \Gamma_{\text{inv}}$ using the couplings to the weak eigenbasis fermion fields, as written in Eq. \ref{Znunu}, because the mixing between the weak and mass eigenbases is unitary. 
We see from Eq.~\ref{Znunu} that the $Z$ boson coupling to electron
  neutrinos is unchanged from the SM, the coupling to muon neutrinos
  $g_{{\nu_L'}_\mu}$ is
  enhanced whereas the coupling to tauon neutrinos
  $g_{{\nu_L'}_\tau}$ is diminished.
  The partial width for each decay $Z \rightarrow \overline{{\nu_L'}_i}
  {\nu_L'}_i$ is $\Gamma_{\nu_i}=|g_{{\nu_i}'}|^2M_Z/(24 \pi) $. 
  There is a partial cancellation between the muon neutrino and the
  tauon neutrino contributions. 
  Working to first order in $\sin \alpha_z \ll 1$ and substituting for it
  using Eq.~\ref{mixing}, we find that the prediction in the DTFHMeg is
  \begin{equation}
    \frac{\Delta \Gamma_{\text{inv}}^{\text{DTFHM}}}{M_Z} =-
    \frac{g_F^2}{48 \pi}
    \left( \frac{M_Z}{M_{Z^\prime}} \right)^2. 
  \end{equation}
  Comparing this to Eq.~\ref{dgamexp}, we find that the DTFHM prediction for the sign
  is in accordance with the data and may fit the inferred invisible
  width of the $Z$ boson some $1.7\sigma$ {\em better}\/ than the SM.

Applying Eq.~\ref{dgamexp} as a constraint implies that, to the
    95$\%$ CL,
  \begin{equation}
  g_F  \frac{1 \text{~TeV}}{M_Z^{\prime}} < 1.05,
  \end{equation}
  which is again satisfied by the whole region of parameter space that fits
  the NCBAs in Eq.~\ref{gm}.

\subsection{Direct $Z^\prime$ search constraints on parameter space}

ATLAS has  released 13 TeV
36.1~fb$^{-1}$  $Z^\prime \rightarrow t \bar t$
searches~\cite{Aaboud:2018mjh,Aaboud:2019roo}, which impose $\sigma \times BR(Z^\prime \rightarrow t \bar t)<10$
fb for large $M_{Z^\prime}$. There is also a search~\cite{Aad:2015osa} for
$Z^\prime 
\rightarrow \tau^+ \tau^-$ for 10 fb$^{-1}$ of 8 TeV data, which imposes
$\sigma \times BR(Z^\prime \rightarrow \tau^+ \tau^-)<3$ fb for large
$M_{Z^\prime}$. These searches constrain the DTFHMeg, but they produce less stringent
constraints than an ATLAS search for $Z^\prime
\rightarrow \mu^+\mu^-$ in 139
fb$^{-1}$ of 13 TeV $pp$ 
collisions~\cite{Aad:2019fac}. We shall therefore concentrate upon this
search. The constraint is in the form of upper limits
upon the fiducial cross-section $\sigma$ times branching ratio
to di-muons $BR(Z^\prime \rightarrow \mu^+ \mu^-)$
as a function of $M_{Z^\prime}$. 
At large $M_{Z^\prime} \approx 6$ TeV, $\sigma \times BR(Z^\prime
\rightarrow \mu^+\mu^-)<0.015$ fb~\cite{atlasData}, and indeed this will prove
to be the most stringent 
$Z^\prime$ direct search constraint, being stronger than the others mentioned
above.

In its recent $Z^\prime \rightarrow \mu^+ \mu^-$ search, ATLAS
defines~\cite{Aad:2019fac} a 
fiducial cross-section $\sigma$ where each muon has transverse momentum $p_T>30$ GeV
and pseudo-rapidity $|\eta|<2.5$, and the di-muon invariant mass satisfies $m_{\mu \mu}>225$
GeV. No evidence for a significant bump in $m_{\mu\mu}$ was found, and so
95$\%$ upper limits on $\sigma \times BR(\mu^+ \mu^-)$ were placed.
Re-casting constraints from such a bump-hunt is
fairly simple: one must simply calculate $\sigma \times BR(\mu^+ \mu^-)$ for the
model in question and apply the bound at the relevant value of $M_{Z^\prime}$
and $\Gamma/M_{Z^\prime}$. Efficiencies are taken into account in the
experimental bound and so there is no need for us to perform a detector
simulation. 
Following Ref.~\cite{Allanach:2019mfl}, for generic $z
\equiv\Gamma/M_{Z^\prime}$, we 
interpolate/extrapolate the upper bound $s(z, M_{Z^\prime})$ on $\sigma \times
BR(\mu^+ \mu^-)$ from those given by ATLAS at $z=0$ and $z=0.1$. In practice, we use a linear
interpolation in $\ln s$:
\begin{equation}
  s(z,M_{Z^\prime}) = s(0,M_{Z^\prime})
  \left[ \frac{s(0.1,M_{Z^\prime})}{s(0,M_{Z^\prime})}\right]^{\frac{z}{0.1}}.  \label{lims}
  \end{equation}
Eq.~\ref{lims} is a reasonable fit~\cite{Allanach:2019mfl} within the range 
$\Gamma/M_{Z^\prime} \in [0,0.1]$. We shall also use Eq.~\ref{lims} to
extrapolate out of this range.

\begin{figure}
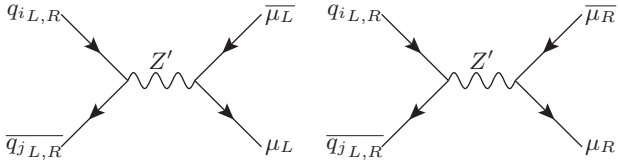

\begin{center}
\begin{axopicture}(170,50)(-40,0)
\Line[arrow](-40,50)(-15,25)
\Line[arrow](-15,25)(-40,0)
\Line[arrow](35,50)(10,25)
\Line[arrow](10,25)(35,0)
\Photon(-15,25)(10,25){3}{3}
\Text(-2.5,33)[c]{$Z^\prime$}
\Text(-50,50)[c]{${q_i}_{L,R}$}
\Text(-50,0)[c]{$\overline{{q_j}_{L,R}}$}
\Text(42,50)[c]{$\overline{\mu_L}$}
\Text(42,0)[c]{$\mu_L$}
\Line[arrow](80,50)(105,25)
\Line[arrow](105,25)(80,0)
\Line[arrow](155,50)(130,25)
\Line[arrow](130,25)(155,0)
\Photon(105,25)(130,25){3}{3}
\Text(117.5,33)[c]{$Z^\prime$}
\Text(70,50)[c]{${q_i}_{L,R}$}
\Text(70,0)[c]{$\overline{{q_j}_{L,R}}$}
\Text(162,50)[c]{$\overline{\mu_R}$}
\Text(162,0)[c]{$\mu_R$}
\end{axopicture}
\end{center}
\caption{Feynman diagrams of tree-level $Z^\prime$ production in the
  LHC by the DTFHM followed by decay into muons, where $q_{i,j} \in \{ u,c,d,s,b \}$ are such that the 
  combination ${q_i}_L \overline{{q_j}_L}$ or
${q_i}_R \overline{{q_j}_R}$
  has zero electric
  charge. In the DTFHMeg, by far the dominant production mode is by
  $q_i=b_L$ and $q_j=b_L$.\label{fig:prod}}    
\end{figure}

In order to use Eq.~\ref{lims}, we must calculate $\sigma \times
BR(\mu^+\mu^-)$, and so we now detail the method of our calculation. 
For the DTFHMeg, we made a {\tt UFO} file\footnote{The {\tt UFO} file is 
  included in the ancillary information submitted with the {\tt arXiv} version
of this paper.} by using
\texttt{FeynRules}~\cite{Degrande:2011ua,Alloul:2013bka}.
We use the {\tt MadGraph_2_6_5}~\cite{Alwall:2014hca} event generator to
estimate $\sigma \times BR(Z^\prime \rightarrow \mu^+ \mu^-)$ for
the
tree-level production processes shown in Fig.~\ref{fig:prod}, in 13 TeV centre of mass energy
$pp$ collisions.
Five flavour parton distribution
functions are used
in order to re-sum the logarithms associated with the initial state
$b$-quark~\cite{Lim:2016wjo}. 

\begin{figure}[ht]
\begin{center}
\unitlength=\textwidth
\includegraphics[width=0.58 \textwidth]{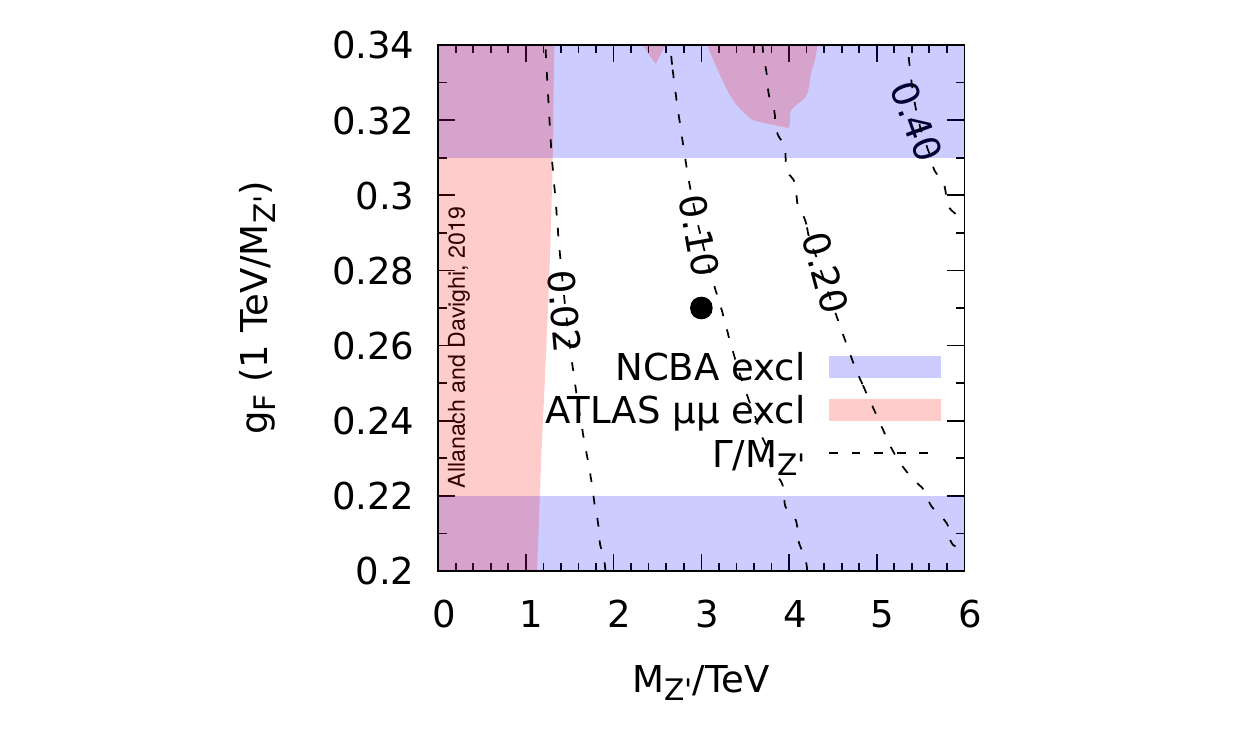}
\caption{\label{fig:const} Constraints on parameter space of the DTFHMeg. The white region is allowed at $95\%$ CL\@.
We show the region excluded at the 95$\%$ CL by the fit to NCBAs, as well
as the 95$\%$ region excluded by the 13 TeV LHC 139 fb$^{-1}$ ATLAS
search~\cite{Aad:2019fac,atlasData}
for $Z^\prime \rightarrow \mu^+ \mu^-$ (labelled by `ATLAS $\mu\mu$ excl'). Other constraints, such as from
$B_s$ mixing, or lepton flavour universality of the $Z$ boson's coupling,
are dealt with in the text and are less restrictive than those shown. The
example point displayed in Table~\ref{tab:egpoint} is shown by the
dot. Values of $\Gamma/M_{Z^\prime}$ label the dashed line contours.}  
\end{center}
\end{figure}
  
\subsection{Combination of constraints}  
  
\begin{figure}[ht]
\begin{center}
\unitlength=\textwidth
\includegraphics[width=0.55 \textwidth]{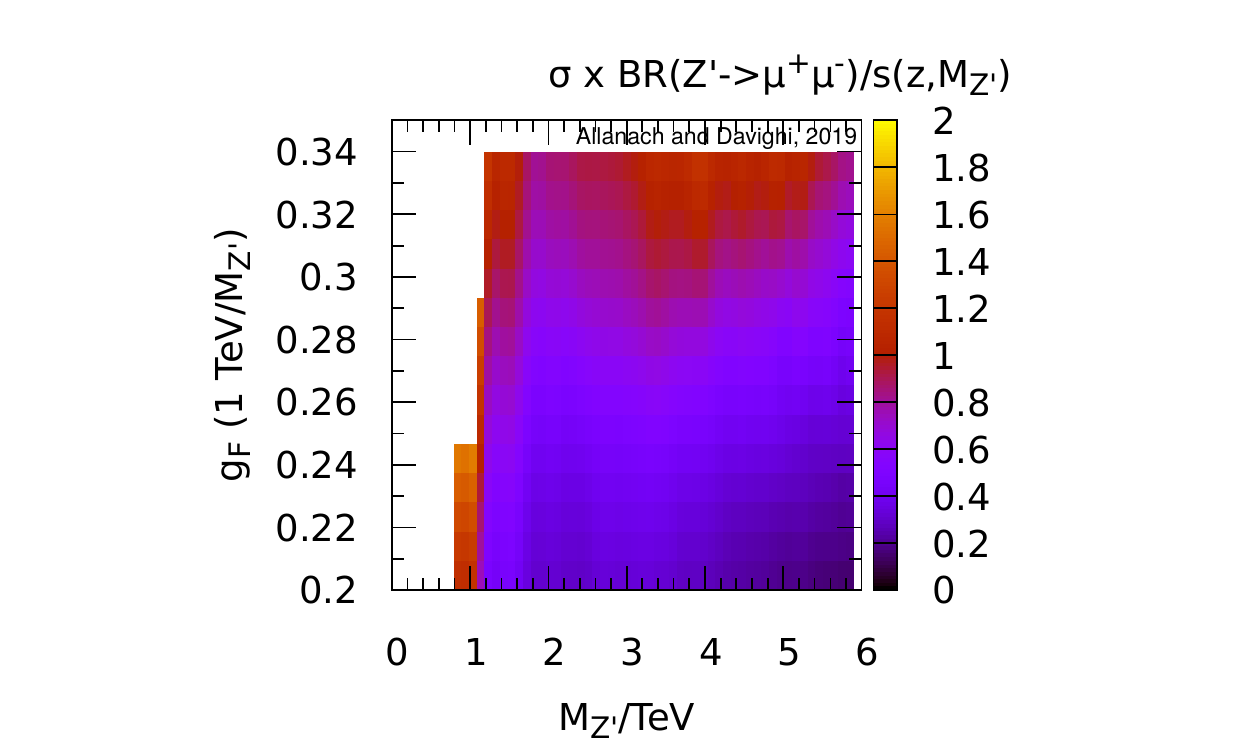}
\end{center}
\caption{\label{fig:sigrel} $\sigma \times BR(Z^\prime \rightarrow \mu^+
\mu^-)$ divided by its upper limit from the ATLAS direct search in the
DTFHMeg. We do not plot any points where this ratio is larger than two,
leading to the white regions on the left-hand side of the figure (which correspond to excluded parameter space).
}  
\end{figure}

We display the resulting constraints upon the DTFHMeg in
Fig.~\ref{fig:const}, with the allowed region shown in white. This allowed region extends
out (beyond the range of the figure) to $M_{Z^\prime}=12.5$ TeV, where the model becomes non-perturbative.
We see that there is plenty of parameter space where the NCBAs are fit and
where current bounds are not contravened. Bounds from $B_s$ mixing and lepton
flavour universality of $Z$ couplings are much weaker than those shown, and do
not impact on the domain of parameter space shown in the figure.
The region to the right-hand side of the $\Gamma/M_{Z^\prime}=0.1$ contour in
the figure is an extrapolation of the bounds in Eq.~\ref{lims}, rather than an
interpolation. We should bear in mind therefore that the extrapolation may be
less accurate the further we move toward the right, away from this contour. 
The branching ratio of $Z^\prime \rightarrow \mu^+ \mu^-$ is approximately
constant over the parameter space shown for $M_{Z^\prime} > 1.2$ TeV. The shape of the
excluded region then depends purely on $\sigma$, which happens to be close to
the inferred search bound for the top excluded region around $M_{Z^\prime}
\sim 3-4$ TeV. This is illustrated by Fig.~\ref{fig:sigrel}.

Note that since the vertical axis is $\propto g_F/ M_{Z^\prime}$, and  since $\sigma
\propto g_F^4$,  $\sigma \times BR(Z^\prime \rightarrow
\mu^+ \mu^-)$ is non-monotonic with respect to $M_{Z^\prime}$ at constant $g_F/ M_{Z^\prime}$.

\begin{table*}[t]
\begin{center}
\begin{tabular}{|c|c|c|c|c|c|c|}\hline
$\Gamma/M_{Z^\prime}$ & $\sigma$/fb & $BR(Z^\prime \rightarrow \mu^+
\mu^-)$ & $BR(Z^\prime \rightarrow t \bar t)$ & $BR(Z^\prime \rightarrow
b \bar b)$ & $BR(Z^\prime \rightarrow \tau^+ \tau^-)$ & $BR(Z^\prime \rightarrow \nu \bar{\nu}^\prime)$ \\ \hline
0.037 & 0.046 & 0.11 & 0.14 & 0.04 & 0.46 & 0.25 \\
\hline  \end{tabular}
\end{center}
\caption{\label{tab:egpoint} Example point in the DTFHMeg parameter space,
$(g_F,M_{Z^\prime})=(0.81,3\ \mathrm{TeV})$.
We display the fiducial production cross-section times branching ratio into di-muons
as $\sigma$. We include the flavour-blind branching ratio to any pair of neutrinos $\nu \bar{\nu}^\prime$. By far the dominant 13 TeV LHC production mode is
$b \bar b \rightarrow Z^\prime$ (the next largest, $b \bar s + s \bar b
\rightarrow Z^\prime$, yields $\sigma=4.7 \times 10^{-5}$ fb). }
\end{table*}

In Table~\ref{tab:egpoint}, we display branching ratio information for an example parameter space point of the DTFHMeg which fits the NCBAs. The dominant production
mode is via $b \bar b \rightarrow Z^\prime$. We see that decays into top
quark pairs, tauon pairs and bottom quark pairs will also be targets for future
searches for the DTFHMeg.

\section{Summary \label{sec:sum}}

We have presented a model which explains the NCBAs
whilst avoiding current
constraints. The model explains some of the coarse features of the
fermion mass spectrum, namely the hierarchical heaviness of the third family
quarks and the smallness of CKM mixing angles. 
The model was obtained by deforming the TFHM in such a way as to retain its
successes whilst 
remedying an ugly feature. The ugly feature involved strong
assumptions that had to be made concerning the charged lepton Yukawa couplings, that 
were not motivated by the symmetries of the model. The deformed model remedies this by
introducing additional charges, such that the $Z^\prime$ resulting
from spontaneous breaking of an anomaly-free $U(1)_F$ symmetry
couples directly to muons already in the weak eigenbasis (in contrast to the
TFHM, where this had to be obtained by $\mu$-$\tau$ mixing).
Another qualitative difference is in how $V_{ts}$ is generated. 
The
 $\overline{s_L}-b_L$ coupling of the $Z^\prime$, which is necessary to explain
the NCBAs, is produced by left-handed strange-bottom mixing. In the TFHM, this
implied that $V_{ts}$ had to be generated by a {\em cancellation}\/ of $\overline{b_L}-s_L$ mixing and
$\overline{t_L}-c_L$ mixing. In the deformed model, this is no longer
necessarily the
case.

We re-cast the most sensitive LHC Run II direct search constraint,  a
139 fb$^{-1}$
ATLAS search for $Z^\prime \rightarrow \mu^+ \mu^-$, for our
DTHFMeg
model, following Ref.~\cite{Allanach:2019mfl}, where a similar analysis was 
performed for the TFHM (and two simplified models).
The result is shown in Fig.~\ref{fig:const}, which, along with the definition of the
model, is the central result of this paper.
Previously,
in Ref.~\cite{Chivukula:2017qsi}, Run I di-jet and di-lepton resonance  
searches (and early Run II searches) were used to
constrain simple $Z^\prime$ models that fit the NCBAs.
The NCBA data have significantly changed since then, as have the search
bounds. Also, since Ref.~\cite{Chivukula:2017qsi} was before the conception of the
TFHM and the DTFHM, it didn't explicitly constrain their parameter spaces. 

In Refs.~\cite{Allanach:2017bta,Allanach:2018odd}, the sensitivity of future
hadron colliders to $Z^\prime$ models that fit the NCBAs was estimated. A 100
TeV future circular collider
(FCC)~\cite{Mangano:2018mur} would have sensitivity to the whole of  
parameter space for one simplified model (MDM)
and the majority of parameter space for another (MUM).
It will be interesting to calculate the future collider reach for both the
DTFHM and the TFHM,
which we suppose may cover the whole perturbative parameter space of each.

\section*{Acknowledgements}

This work has been partially supported by STFC consolidated grant
ST/P000681/1. We thank Henning Bahl and the Cambridge Pheno Working group for helpful
discussions. JD has been supported by The Cambridge Trust.

\appendix

\section{Froggatt-Nielsen structure to obtain a diagonal up Yukawa
  matrix \label{sec:app}}

In the example case that we chose to study, denoted the DTFHMeg, we made some specific choices for the mixing matrices $\{V_P\}$. In particular, we assumed $V_{u_L}=1$ and $V_{u_R}=1$ for simplicity, with the CKM mixing arising solely from the down-type quarks. This `up-alignment' requires there be a sensible limit in which the effective up-type Yukawa matrix $Y_u$ is approximately diagonal. In this Appendix, we sketch
how this might be obtained from a more detailed model utilising the Froggatt-Nielsen~\cite{Froggatt:1978nt} mechanism. Evidently, such a model must ultimately break the $U(2)$ flavour symmetry acting on the light up-type quarks, in order to suppress the $(Y_u)_{12}$ and $(Y_u)_{21}$ matrix elements with respect to their diagonal counterparts.
There are presumably other ways of model building such detailed Yukawa structures, but our purpose here is only to provide an existence proof of such mechanisms, with more detailed model building being well beyond the scope of the present paper.

\begin{figure}
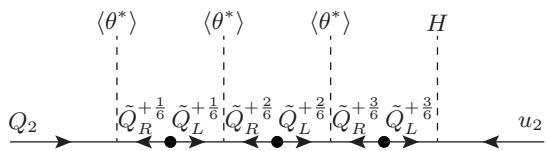

\begin{center}
\begin{axopicture}(200,50)(0,0)
  \Line[arrow](0,0)(40,0)
  \Text(5,7)[c]{$Q_2$}
  \Line[dash](40,0)(40,40)
  \Text(40,45)[c]{$\langle \theta^* \rangle$}
  \Line[arrow](60,0)(40,0)
  \GCirc(60,0){2}{0}
  \Line[arrow](60,0)(80,0)
  \Text(50,10)[c]{$\tilde Q_{R}^{+\frac{1}{6}}$}
  \Text(70,10)[c]{$\tilde Q_{L}^{+\frac{1}{6}}$}
  \Line[dash](80,0)(80,40)
  \Text(80,45)[c]{$\langle \theta^* \rangle$}  
  \Line[arrow](100,0)(80,0)
  \GCirc(100,0){2}{0}
  \Line[arrow](100,0)(120,0)
  \Text(90,10)[c]{$\tilde Q_{R}^{+\frac{2}{6}}$}
  \Text(110,10)[c]{$\tilde Q_{L}^{+\frac{2}{6}}$}
  \Line[dash](120,0)(120,40)
  \Text(120,45)[c]{$\langle \theta^* \rangle$}  
  \Line[arrow](140,0)(120,0)
  \GCirc(140,0){2}{0}
  \Line[arrow](140,0)(160,0)
  \Text(130,10)[c]{$\tilde Q_{R}^{+\frac{3}{6}}$}
  \Text(150,10)[c]{$\tilde Q_{L}^{+\frac{3}{6}}$}
  \Line[dash](160,0)(160,40)
  \Text(160,45)[c]{$H$}
  \Line[arrow](200,0)(160,0)
    \Text(195,7)[c]{$u_2$}
\end{axopicture}
\end{center}
\caption{Feynman diagram yielding an effective non-zero entry in
  $(Y_u)_{22}$. Each solid blob represents a mass insertion factor of {$1/M$}. \label{fig:mc}}
\end{figure}

Froggatt-Nielsen models postulate the existence of heavy vector-like fermions in the same
representations as the SM chiral fermions, but with independent charges under $U(1)_F$. We shall here denote such a heavy fermion by its SM counterpart field but with a tilde, {\em i.e.}\ $\tilde Q^{F}_{L,R} \sim (3, 2, 1/6, F)$, 
$\tilde L^{F}_{L,R} \sim (1, 2, -1/2, F)$, 
$\tilde e^{F}_{L,R} \sim (1, 1, -1, F)$, 
$\tilde d^{F}_{L,R} \sim (3, 1, -1/3, F)$, 
$\tilde u^{F}_{L,R} \sim (3, 1, 2/3, F)$.
The idea is that their masses (denoted loosely and collectively as $M$) are
larger than $v_F$, say five times larger or 
so. Then, after $U(1)_F$ breaking and integrating out the heavy fermions, the model generates $U(1)_F$-violating operators  in the SM effective field theory,  such as that depicted by the Feynman diagram in
Fig.~\ref{fig:mc}. Each such operator is suppressed by a power of $v_F/M$
that is set by the total $U(1)_F$ charge of the  Yukawa operator in the SM effective field theory. 

Hereafter we assume that $F_\theta=1/6$.
The coefficient of the effective operator in Fig. \ref{fig:mc} would be equal to the
  product of four dimensionless coupling constants and a factor of $(v_F/M)^3$. The typical assumption in Froggatt-Nielsen models is then that the fundamental dimensionless coupling constants are all roughly equal to 1, and
  that all the heavy fermion fields have roughly the same mass, denoted by $M$, which would yield a prediction
  $(Y_u)_{22}\sim \mathcal{O}(v_F/M)^3$. It is by changing these strong
  assumptions that one may 
  generate operators which are {\em further} suppressed for the off-diagonal entries of
  $Y_u$, and thus predict $V_{u_L}\approx V_{u_R} \approx 1$ as we assumed in the DTFHMeg. Such
  suppression may come 
  from a suppression of some of the fundamental dimensionless coupling constants, which
  may be set by other symmetries or dynamics of the model.

To give an explicit realisation of this idea that is pertinent to the DTFHMeg, suppose that
  the effective coupling $(Y_u)_{22}$ is mediated by a `spaghetti diagram' involving the $\tilde Q_{L,R}$ fields, as in Fig.~\ref{fig:mc}, whereas the effective coupling $(Y_u)_{11}$ is mediated by the $\tilde u_{L, R}$
  fields, as in Fig.~\ref{fig:mu}. This could be achieved if the fundamental dimensionless couplings of
  $\overline{u_1} H \tilde Q^{+3}_L $ and $\overline{\tilde u^{+3}_R} H  Q_2$ were zero or approximately zero for some
  reason (thus explicitly breaking the $U(2)$ flavour symmetry that acts on light up-type quarks), but all other gauge invariant Yukawa couplings were of the same order. Then a mixing term such as $(Y_u)_{12}$
  must be 
  mediated by {\em both}\/ $\tilde u_{L, R}$ {\em and}\/ 
$\tilde Q_{L,R}$ fields, leading to an
  additional suppression because the dimensionality of the operator is higher
  than the diagonal entries. 
   Thus, the off-diagonal effective couplings $(Y_u)_{12}$ and $(Y_u)_{21}$ would be heavily suppressed compared to
  $(Y_u)_{11}$ or $(Y_u)_{22}$.\footnote{Clearly this example is not a complete model as we have sketched it here, because it predicts a charm mass of order the up mass. More details of the model can be fleshed out to
    remedy this. For example, by making the $\tilde u_{L,R}$ fields heavier than the
    $\tilde Q_{L.R}$ fields (by a factor of order $(m_c/m_u)^{1/3}\sim 10$), or assuming that they have somewhat smaller Yukawa couplings with $\theta$ (by a factor of order $(m_u/m_c)^{1/3}\sim 1/10$), one can explain the hierarchy between the up and charm masses without any large hierarchies in the fundamental parameters. In any case, our purpose here is merely to show the kind of considerations one can make within a Froggatt-Nielsen framework in order to
    suppress certain operators, not build a complete model of fermion masses.} Similar methods can be employed to suppress
  other off-diagonal terms in the up-quark or charged lepton sectors, as required by the particular example case being explored. 

\begin{figure}
\begin{center}
\begin{axopicture}(200,50)(0,0)
  \Line[arrow](0,0)(40,0)
  \Text(5,7)[c]{$Q_1$}
  \Line[dash](40,0)(40,40)
  \Text(40,45)[c]{$H$}
  \Line[arrow](60,0)(40,0)
  \GCirc(60,0){2}{0}
  \Line[arrow](60,0)(80,0)
  \Text(50,10)[c]{$\tilde u_{R}^{+\frac{3}{6}}$}
  \Text(70,10)[c]{$\tilde u_{L}^{+\frac{3}{6}}$}
  \Line[dash](80,0)(80,40)
  \Text(80,45)[c]{$\langle \theta \rangle$}  
  \Line[arrow](100,0)(80,0)
  \GCirc(100,0){2}{0}
  \Line[arrow](100,0)(120,0)
  \Text(90,10)[c]{$\tilde u_{R}^{+\frac{2}{6}}$}
  \Text(110,10)[c]{$\tilde u_{L}^{+\frac{2}{6}}$}
  \Line[dash](120,0)(120,40)
  \Text(120,45)[c]{$\langle \theta \rangle$}  
  \Line[arrow](140,0)(120,0)
  \GCirc(140,0){2}{0}
  \Line[arrow](140,0)(160,0)
  \Text(130,10)[c]{$\tilde u_{R}^{+\frac{1}{6}}$}
  \Text(150,10)[c]{$\tilde u_{L}^{+\frac{1}{6}}$}
  \Line[dash](160,0)(160,40)
  \Text(160,45)[c]{$\langle \theta \rangle $}
  \Line[arrow](200,0)(160,0)
    \Text(195,7)[c]{$u_1$}
\end{axopicture}
\end{center}
\caption{Feynman diagram yielding an effective non-zero entry in
  $(Y_u)_{11}$. Each solid blob represents a mass insertion of {$1/M$}. \label{fig:mu}}
\end{figure}

\bibliographystyle{JHEP-2}
\bibliography{beyond_TFHM}

\newcommand{\noop}[1]{}
\providecommand{\href}[2]{#2}\begingroup\raggedright\begin{thebibliography}{10}

\bibitem{Aaij:2017vbb}
{\bf LHCb} Collaboration, R.~Aaij {\em et.~al.}, {\it {Test of lepton
  universality with $B^{0} \rightarrow K^{*0}\ell^{+}\ell^{-}$ decays}},  {\em
  JHEP} {\bf 08} (2017) 055 [\href{http://arXiv.org/abs/1705.05802}{{\tt
  1705.05802}}].

\bibitem{CERN-EP-2019-043}
{\bf LHCb Collaboration} Collaboration, {\it {Search for lepton-universality
  violation in $B^+\to K^+\ell^+\ell^-$ decays}},  Tech. Rep. CERN-EP-2019-043.
  LHCB-PAPER-2019-009, CERN, Geneva, Mar, 2019.

\bibitem{Aaboud:2018mst}
{\bf ATLAS} Collaboration, M.~Aaboud {\em et.~al.}, {\it {Study of the rare
  decays of $B^0_s$ and $B^0$ mesons into muon pairs using data collected
  during 2015 and 2016 with the ATLAS detector}},  {\em JHEP} {\bf 04} (2019)
  098 [\href{http://arXiv.org/abs/1812.03017}{{\tt 1812.03017}}].

\bibitem{Chatrchyan:2013bka}
{\bf CMS} Collaboration, S.~Chatrchyan {\em et.~al.}, {\it {Measurement of the
  $B^0_s \to \mu^+ \mu^-$ Branching Fraction and Search for $B^0 \to \mu^+
  \mu^-$ with the CMS Experiment}},  {\em Phys. Rev. Lett.} {\bf 111} (2013)
  101804 [\href{http://arXiv.org/abs/1307.5025}{{\tt 1307.5025}}].

\bibitem{CMS:2014xfa}
{\bf CMS, LHCb} Collaboration, V.~Khachatryan {\em et.~al.}, {\it {Observation
  of the rare $B^0_s\to\mu^+\mu^-$ decay from the combined analysis of CMS and
  LHCb data}},  {\em Nature} {\bf 522} (2015) 68--72
  [\href{http://arXiv.org/abs/1411.4413}{{\tt 1411.4413}}].

\bibitem{Aaij:2017vad}
{\bf LHCb} Collaboration, R.~Aaij {\em et.~al.}, {\it {Measurement of the
  $B^0_s\to\mu^+\mu^-$ branching fraction and effective lifetime and search for
  $B^0\to\mu^+\mu^-$ decays}},  {\em Phys. Rev. Lett.} {\bf 118} (2017), no.~19
  191801 [\href{http://arXiv.org/abs/1703.05747}{{\tt 1703.05747}}].

\bibitem{Aaij:2013qta}
{\bf LHCb} Collaboration, R.~Aaij {\em et.~al.}, {\it {Measurement of
  Form-Factor-Independent Observables in the Decay $B^{0} \to K^{*0} \mu^+
  \mu^-$}},  {\em Phys. Rev. Lett.} {\bf 111} (2013) 191801
  [\href{http://arXiv.org/abs/1308.1707}{{\tt 1308.1707}}].

\bibitem{Aaij:2015oid}
{\bf LHCb} Collaboration, R.~Aaij {\em et.~al.}, {\it {Angular analysis of the
  $B^{0} \to K^{*0} \mu^{+} \mu^{-}$ decay using 3 fb$^{-1}$ of integrated
  luminosity}},  {\em JHEP} {\bf 02} (2016) 104
  [\href{http://arXiv.org/abs/1512.04442}{{\tt 1512.04442}}].

\bibitem{ATLAS-CONF-2017-023}
{\bf ATLAS Collaboration} Collaboration, {\it {Angular analysis of $B^0_d \to
  K^{*}\mu^+\mu^-$ decays in $pp$ collisions at $\sqrt{s}= 8$ TeV with the
  ATLAS detector}},  Tech. Rep. ATLAS-CONF-2017-023, CERN, Geneva, Apr, 2017.

\bibitem{CMS-PAS-BPH-15-008}
{\bf CMS Collaboration} Collaboration, {\it {Measurement of the $P_1$ and
  $P_5'$ angular parameters of the decay $\mathrm{B}^0 \to \mathrm{K}^{*0}
  \mu^+ \mu^-$ in proton-proton collisions at $\sqrt{s}=8~\mathrm{TeV}$}},
  Tech. Rep. CMS-PAS-BPH-15-008, CERN, Geneva, 2017.

\bibitem{Khachatryan:2015isa}
{\bf CMS} Collaboration, V.~Khachatryan {\em et.~al.}, {\it {Angular analysis
  of the decay $B^0 \to K^{*0} \mu^+ \mu^-$ from pp collisions at $\sqrt s = 8$
  TeV}},  {\em Phys. Lett.} {\bf B753} (2016) 424--448
  [\href{http://arXiv.org/abs/1507.08126}{{\tt 1507.08126}}].

\bibitem{Bobeth:2017vxj}
C.~Bobeth, M.~Chrzaszcz, D.~van Dyk and J.~Virto, {\it {Long-distance effects
  in $B\rightarrow K^*\ell \ell $ from analyticity}},  {\em Eur. Phys. J.} {\bf
  C78} (2018), no.~6 451 [\href{http://arXiv.org/abs/1707.07305}{{\tt
  1707.07305}}].

\bibitem{DAmico:2017mtc}
G.~D'Amico, M.~Nardecchia, P.~Panci, F.~Sannino, A.~Strumia, R.~Torre and
  A.~Urbano, {\it {Flavour anomalies after the $R_{K^*}$ measurement}},  {\em
  JHEP} {\bf 09} (2017) 010 [\href{http://arXiv.org/abs/1704.05438}{{\tt
  1704.05438}}].

\bibitem{Alguero:2019ptt}
M.~Algueró, B.~Capdevila, A.~Crivellin, S.~Descotes-Genon, P.~Masjuan,
  J.~Matias and J.~Virto, {\it {Addendum: "Patterns of New Physics in $b\to s
  \ell^+\ell^-$ transitions in the light of recent data" and "Are we
  overlooking Lepton Flavour Universal New Physics in $b \to s \ell\ell\,$?"}},
   \href{http://arXiv.org/abs/1903.09578}{{\tt 1903.09578}}.

\bibitem{Alok:2019ufo}
A.~K. Alok, A.~Dighe, S.~Gangal and D.~Kumar, {\it {Continuing search for new
  physics in $b \to s \mu \mu$ decays: two operators at a time}},
  \href{http://arXiv.org/abs/1903.09617}{{\tt 1903.09617}}.

\bibitem{Ciuchini:2019usw}
M.~Ciuchini, A.~M. Coutinho, M.~Fedele, E.~Franco, A.~Paul, L.~Silvestrini and
  M.~Valli, {\it {New Physics in $b \to s \ell^+ \ell^-$ confronts new data on
  Lepton Universality}},  \href{http://arXiv.org/abs/1903.09632}{{\tt
  1903.09632}}.

\bibitem{Aebischer:2019mlg}
J.~Aebischer, W.~Altmannshofer, D.~Guadagnoli, M.~Reboud, P.~Stangl and D.~M.
  Straub, {\it {$B$-decay discrepancies after Moriond 2019}},
  \href{http://arXiv.org/abs/1903.10434}{{\tt 1903.10434}}.

\bibitem{Kowalska:2019ley}
K.~Kowalska, D.~Kumar and E.~M. Sessolo, {\it {Implications for New Physics in
  $b\to s \mu\mu$ transitions after recent measurements by Belle and LHCb}},
  \href{http://arXiv.org/abs/1903.10932}{{\tt 1903.10932}}.

\bibitem{Arbey:2019duh}
A.~Arbey, T.~Hurth, F.~Mahmoudi, D.~Martinez~Santos and S.~Neshatpour, {\it
  {Update on the b->s anomalies}},  \href{http://arXiv.org/abs/1904.08399}{{\tt
  1904.08399}}.

\bibitem{Ellis:2017nrp}
J.~Ellis, M.~Fairbairn and P.~Tunney, {\it {Anomaly-Free Models for Flavour
  Anomalies}},  {\em Eur. Phys. J.} {\bf C78} (2018), no.~3 238
  [\href{http://arXiv.org/abs/1705.03447}{{\tt 1705.03447}}].

\bibitem{Allanach:2018vjg}
B.~C. Allanach, J.~Davighi and S.~Melville, {\it {An Anomaly-free Atlas:
  charting the space of flavour-dependent gauged $U(1)$ extensions of the
  Standard Model}},  {\em JHEP} {\bf 02} (2019) 082
  [\href{http://arXiv.org/abs/1812.04602}{{\tt 1812.04602}}].

\bibitem{Gauld:2013qba}
R.~Gauld, F.~Goertz and U.~Haisch, {\it {On minimal $Z'$ explanations of the
  $B\to K^*\mu^+\mu^-$ anomaly}},  {\em Phys. Rev.} {\bf D89} (2014) 015005
  [\href{http://arXiv.org/abs/1308.1959}{{\tt 1308.1959}}].

\bibitem{Buras:2013dea}
A.~J. Buras, F.~De~Fazio and J.~Girrbach, {\it {331 models facing new $b \to
  s\mu^+ \mu^-$ data}},  {\em JHEP} {\bf 02} (2014) 112
  [\href{http://arXiv.org/abs/1311.6729}{{\tt 1311.6729}}].

\bibitem{Buras:2013qja}
A.~J. Buras and J.~Girrbach, {\it {Left-handed $Z'$ and $Z$ FCNC quark
  couplings facing new $b \to s \mu^+ \mu^-$ data}},  {\em JHEP} {\bf 12}
  (2013) 009 [\href{http://arXiv.org/abs/1309.2466}{{\tt 1309.2466}}].

\bibitem{Altmannshofer:2014cfa}
W.~Altmannshofer, S.~Gori, M.~Pospelov and I.~Yavin, {\it {Quark flavor
  transitions in $L_\mu-L_\tau$ models}},  {\em Phys. Rev.} {\bf D89} (2014)
  095033 [\href{http://arXiv.org/abs/1403.1269}{{\tt 1403.1269}}].

\bibitem{Buras:2014yna}
A.~J. Buras, F.~De~Fazio and J.~Girrbach-Noe, {\it {$Z$-$Z'$ mixing and
  $Z$-mediated FCNCs in $SU(3)_{C} \times SU(3)_{L} \times U(1)_{X}$ models}},
  {\em JHEP} {\bf 08} (2014) 039 [\href{http://arXiv.org/abs/1405.3850}{{\tt
  1405.3850}}].

\bibitem{Crivellin:2015mga}
A.~Crivellin, G.~D'Ambrosio and J.~Heeck, {\it {Explaining
  $h\to\mu^\pm\tau^\mp$, $B\to K^* \mu^+\mu^-$ and $B\to K \mu^+\mu^-/B\to K
  e^+e^-$ in a two-Higgs-doublet model with gauged $L_\mu-L_\tau$}},  {\em
  Phys. Rev. Lett.} {\bf 114} (2015) 151801
  [\href{http://arXiv.org/abs/1501.00993}{{\tt 1501.00993}}].

\bibitem{Crivellin:2015lwa}
A.~Crivellin, G.~D'Ambrosio and J.~Heeck, {\it {Addressing the LHC flavor
  anomalies with horizontal gauge symmetries}},  {\em Phys. Rev.} {\bf D91}
  (2015), no.~7 075006 [\href{http://arXiv.org/abs/1503.03477}{{\tt
  1503.03477}}].

\bibitem{Sierra:2015fma}
D.~Aristizabal~Sierra, F.~Staub and A.~Vicente, {\it {Shedding light on the
  $b\to s$ anomalies with a dark sector}},  {\em Phys. Rev.} {\bf D92} (2015),
  no.~1 015001 [\href{http://arXiv.org/abs/1503.06077}{{\tt 1503.06077}}].

\bibitem{Crivellin:2015era}
A.~Crivellin, L.~Hofer, J.~Matias, U.~Nierste, S.~Pokorski and J.~Rosiek, {\it
  {Lepton-flavour violating $B$ decays in generic $Z'$ models}},  {\em Phys.
  Rev.} {\bf D92} (2015), no.~5 054013
  [\href{http://arXiv.org/abs/1504.07928}{{\tt 1504.07928}}].

\bibitem{Celis:2015ara}
A.~Celis, J.~Fuentes-Martin, M.~Jung and H.~Serodio, {\it {Family nonuniversal
  Z′ models with protected flavor-changing interactions}},  {\em Phys. Rev.}
  {\bf D92} (2015), no.~1 015007 [\href{http://arXiv.org/abs/1505.03079}{{\tt
  1505.03079}}].

\bibitem{Greljo:2015mma}
A.~Greljo, G.~Isidori and D.~Marzocca, {\it {On the breaking of Lepton Flavor
  Universality in B decays}},  {\em JHEP} {\bf 07} (2015) 142
  [\href{http://arXiv.org/abs/1506.01705}{{\tt 1506.01705}}].

\bibitem{Altmannshofer:2015mqa}
W.~Altmannshofer and I.~Yavin, {\it {Predictions for lepton flavor universality
  violation in rare B decays in models with gauged $L_\mu - L_\tau$}},  {\em
  Phys. Rev.} {\bf D92} (2015), no.~7 075022
  [\href{http://arXiv.org/abs/1508.07009}{{\tt 1508.07009}}].

\bibitem{Allanach:2015gkd}
B.~Allanach, F.~S. Queiroz, A.~Strumia and S.~Sun, {\it {$Z^\prime$ models for
  the LHCb and $g-2$ muon anomalies}},  {\em Phys. Rev.} {\bf D93} (2016),
  no.~5 055045 [\href{http://arXiv.org/abs/1511.07447}{{\tt 1511.07447}}].
  [Erratum: Phys. Rev.D95,no.11,119902(2017)].

\bibitem{Falkowski:2015zwa}
A.~Falkowski, M.~Nardecchia and R.~Ziegler, {\it {Lepton Flavor
  Non-Universality in B-meson Decays from a U(2) Flavor Model}},  {\em JHEP}
  {\bf 11} (2015) 173 [\href{http://arXiv.org/abs/1509.01249}{{\tt
  1509.01249}}].

\bibitem{Chiang:2016qov}
C.-W. Chiang, X.-G. He and G.~Valencia, {\it {Z′ model for
  b→sℓ$\overline{ℓ}$ flavor anomalies}},  {\em Phys. Rev.} {\bf D93}
  (2016), no.~7 074003 [\href{http://arXiv.org/abs/1601.07328}{{\tt
  1601.07328}}].

\bibitem{Becirevic:2016zri}
D.~Bečirević, O.~Sumensari and R.~Zukanovich~Funchal, {\it {Lepton flavor
  violation in exclusive $b\rightarrow s$ decays}},  {\em Eur. Phys. J.} {\bf
  C76} (2016), no.~3 134 [\href{http://arXiv.org/abs/1602.00881}{{\tt
  1602.00881}}].

\bibitem{Boucenna:2016wpr}
S.~M. Boucenna, A.~Celis, J.~Fuentes-Martin, A.~Vicente and J.~Virto, {\it
  {Non-abelian gauge extensions for B-decay anomalies}},  {\em Phys. Lett.}
  {\bf B760} (2016) 214--219 [\href{http://arXiv.org/abs/1604.03088}{{\tt
  1604.03088}}].

\bibitem{Boucenna:2016qad}
S.~M. Boucenna, A.~Celis, J.~Fuentes-Martin, A.~Vicente and J.~Virto, {\it
  {Phenomenology of an $SU(2) \times SU(2) \times U(1)$ model with
  lepton-flavour non-universality}},  {\em JHEP} {\bf 12} (2016) 059
  [\href{http://arXiv.org/abs/1608.01349}{{\tt 1608.01349}}].

\bibitem{Ko:2017lzd}
P.~Ko, Y.~Omura, Y.~Shigekami and C.~Yu, {\it {LHCb anomaly and B physics in
  flavored Z′ models with flavored Higgs doublets}},  {\em Phys. Rev.} {\bf
  D95} (2017), no.~11 115040 [\href{http://arXiv.org/abs/1702.08666}{{\tt
  1702.08666}}].

\bibitem{Alonso:2017bff}
R.~Alonso, P.~Cox, C.~Han and T.~T. Yanagida, {\it {Anomaly-free local
  horizontal symmetry and anomaly-full rare B-decays}},  {\em Phys. Rev.} {\bf
  D96} (2017), no.~7 071701 [\href{http://arXiv.org/abs/1704.08158}{{\tt
  1704.08158}}].

\bibitem{Alonso:2017uky}
R.~Alonso, P.~Cox, C.~Han and T.~T. Yanagida, {\it {Flavoured $B−L$ local
  symmetry and anomalous rare $B$ decays}},  {\em Phys. Lett.} {\bf B774}
  (2017) 643--648 [\href{http://arXiv.org/abs/1705.03858}{{\tt 1705.03858}}].

\bibitem{1674-1137-42-3-033104}
Y.~Tang and Y.-L. Wu, {\it Flavor non-universal gauge interactions and
  anomalies in b-meson decays},  {\em Chinese Physics C} {\bf 42} (2018), no.~3
  033104.

\bibitem{Bonilla:2017lsq}
C.~Bonilla, T.~Modak, R.~Srivastava and J.~W.~F. Valle, {\it
  {$U(1)_{B_3-3L_\mu}$ gauge symmetry as the simplest description of $b\to s$
  anomalies}},  \href{http://arXiv.org/abs/1705.00915}{{\tt 1705.00915}}.

\bibitem{Bhatia:2017tgo}
D.~Bhatia, S.~Chakraborty and A.~Dighe, {\it {Neutrino mixing and $R_K$ anomaly
  in U(1)$_X$ models: a bottom-up approach}},  {\em JHEP} {\bf 03} (2017) 117
  [\href{http://arXiv.org/abs/1701.05825}{{\tt 1701.05825}}].

\bibitem{CHEN2018420}
C.-H. Chen and T.~Nomura, {\it Penguin b→sℓ′+ℓ′− and b-meson
  anomalies in a gauged lμ−lτ},  {\em Physics Letters B} {\bf 777} (2018)
  420 -- 427.

\bibitem{Faisel:2017glo}
G.~Faisel and J.~Tandean, {\it {Connecting $ b\to s\ell \overline{\ell} $
  anomalies to enhanced rare nonleptonic $ {\overline{B}}_s^0 $ decays in
  $Z′$ model}},  {\em JHEP} {\bf 02} (2018) 074
  [\href{http://arXiv.org/abs/1710.11102}{{\tt 1710.11102}}].

\bibitem{PhysRevD.97.115003}
K.~Fuyuto, H.-L. Li and J.-H. Yu, {\it Implications of hidden gauged
  $u\mathbf{(}1\mathbf{)}$ model for $b$ anomalies},  {\em Phys. Rev. D} {\bf
  97} (Jun, 2018) 115003.

\bibitem{Bian:2017xzg}
L.~Bian, H.~M. Lee and C.~B. Park, {\it {$B$-meson anomalies and Higgs physics
  in flavored $U(1)'$ model}},  {\em Eur. Phys. J.} {\bf C78} (2018), no.~4 306
  [\href{http://arXiv.org/abs/1711.08930}{{\tt 1711.08930}}].

\bibitem{PhysRevD.97.075035}
M.~Abdullah, M.~Dalchenko, B.~Dutta, R.~Eusebi, P.~Huang, T.~Kamon, D.~Rathjens
  and A.~Thompson, {\it Bottom-quark fusion processes at the lhc for probing
  ${Z}^{\ensuremath{'}}$ models and $b$-meson decay anomalies},  {\em Phys.
  Rev. D} {\bf 97} (Apr, 2018) 075035.

\bibitem{King:2018fcg}
S.~F. King, {\it {$R_{K^{(*)}}$ and the origin of Yukawa couplings}},
  \href{http://arXiv.org/abs/1806.06780}{{\tt 1806.06780}}.

\bibitem{Duan:2018akc}
G.~H. Duan, X.~Fan, M.~Frank, C.~Han and J.~M. Yang, {\it {A minimal
  $U(1)^\prime$ extension of MSSM in light of the B decay anomaly}},
  \href{http://arXiv.org/abs/1808.04116}{{\tt 1808.04116}}.

\bibitem{Allanach:2018lvl}
B.~C. Allanach and J.~Davighi, {\it {Third Family Hypercharge Model for
  $R_{K^{(\ast)}}$ and Aspects of the Fermion Mass Problem}},  {\em JHEP} {\bf
  12} (2018) 075 [\href{http://arXiv.org/abs/1809.01158}{{\tt 1809.01158}}].

\bibitem{Allanach:2018odd}
B.~C. Allanach, T.~Corbett, M.~J. Dolan and T.~You, {\it {Hadron Collider
  Sensitivity to Fat Flavourful $Z^\prime$s for $R_{K^{(\ast)}}$}},
  \href{http://arXiv.org/abs/1810.02166}{{\tt 1810.02166}}.

\bibitem{Crivellin:2016ejn}
A.~Crivellin, J.~Fuentes-Martin, A.~Greljo and G.~Isidori, {\it {Lepton Flavor
  Non-Universality in B decays from Dynamical Yukawas}},  {\em Phys. Lett.}
  {\bf B766} (2017) 77--85 [\href{http://arXiv.org/abs/1611.02703}{{\tt
  1611.02703}}].

\bibitem{Kamenik:2017tnu}
J.~F. Kamenik, Y.~Soreq and J.~Zupan, {\it {Lepton flavor universality
  violation without new sources of quark flavor violation}},  {\em Phys. Rev.}
  {\bf D97} (2018), no.~3 035002 [\href{http://arXiv.org/abs/1704.06005}{{\tt
  1704.06005}}].

\bibitem{Camargo-Molina:2018cwu}
J.~E. Camargo-Molina, A.~Celis and D.~A. Faroughy, {\it {Anomalies in Bottom
  from new physics in Top}},  {\em Phys. Lett.} {\bf B784} (2018) 284--293
  [\href{http://arXiv.org/abs/1805.04917}{{\tt 1805.04917}}].

\bibitem{Davighi:2019jwf}
J.~Davighi, {\it {Connecting neutral current $B$ anomalies with the heaviness
  of the third family}},  2019.
\newblock \href{http://arXiv.org/abs/1905.06073}{{\tt 1905.06073}}.

\bibitem{PhysRevD.98.030001}
{\bf Particle Data Group} Collaboration, M.~Tanabashi {\em et.~al.}, {\it
  Review of particle physics},  {\em Phys. Rev. D} {\bf 98} (Aug, 2018) 030001.

\bibitem{b_c_allanach_2018_1478085}
B.~C. Allanach, J.~Davighi and S.~Melville, {\it {Anomaly-free,
  flavour-dependent U(1) charge assignments for Standard Model/Standard Model
  plus three right-handed neutrino fermionic content}},  Dec., 2018.
\newblock file {\tt SMcharges10}.

\bibitem{King:2019lal}
D.~King, A.~Lenz and T.~Rauh, {\it {Bs mixing observables and Vtd/Vts from sum
  rules}},  \href{http://arXiv.org/abs/1904.00940}{{\tt 1904.00940}}.

\bibitem{Amhis:2016xyh}
{\bf HFLAV} Collaboration, Y.~Amhis {\em et.~al.}, {\it {Averages of
  $b$-hadron, $c$-hadron, and $\tau$-lepton properties as of summer 2016}},
  {\em Eur. Phys. J.} {\bf C77} (2017), no.~12 895
  [\href{http://arXiv.org/abs/1612.07233}{{\tt 1612.07233}}]. online update at
  http://www.slac.stanford.edu.

\bibitem{Allanach:2019mfl}
B.~C. Allanach, J.~M. Butterworth and T.~Corbett, {\it {Collider Constraints on
  $Z^\prime$ Models for Neutral Current $B-$Anomalies}},
  \href{http://arXiv.org/abs/1904.10954}{{\tt 1904.10954}}.

\bibitem{DiLuzio:2017fdq}
L.~Di~Luzio, M.~Kirk and A.~Lenz, {\it {Updated $B_s$-mixing constraints on new
  physics models for $b\to s\ell^+\ell^-$ anomalies}},  {\em Phys. Rev.} {\bf
  D97} (2018), no.~9 095035 [\href{http://arXiv.org/abs/1712.06572}{{\tt
  1712.06572}}].

\bibitem{Charles:2013aka}
J.~Charles, S.~Descotes-Genon, Z.~Ligeti, S.~Monteil, M.~Papucci and
  K.~Trabelsi, {\it {Future sensitivity to new physics in $B_d, B_s$, and K
  mixings}},  {\em Phys. Rev.} {\bf D89} (2014), no.~3 033016
  [\href{http://arXiv.org/abs/1309.2293}{{\tt 1309.2293}}].

\bibitem{ALEPH:2005ab}
{\bf ALEPH, DELPHI, L3, OPAL, SLD, LEP Electroweak Working Group, SLD
  Electroweak Group, SLD Heavy Flavour Group} Collaboration, S.~Schael {\em
  et.~al.}, {\it {Precision electroweak measurements on the $Z$ resonance}},
  {\em Phys. Rept.} {\bf 427} (2006) 257--454
  [\href{http://arXiv.org/abs/hep-ex/0509008}{{\tt hep-ex/0509008}}].

\bibitem{Aaboud:2018mjh}
{\bf ATLAS} Collaboration, M.~Aaboud {\em et.~al.}, {\it {Search for heavy
  particles decaying into top-quark pairs using lepton-plus-jets events in
  protonproton collisions at $\sqrt{s} = 13$ $\text {TeV}$ with the ATLAS
  detector}},  {\em Eur. Phys. J.} {\bf C78} (2018), no.~7 565
  [\href{http://arXiv.org/abs/1804.10823}{{\tt 1804.10823}}].

\bibitem{Aaboud:2019roo}
{\bf ATLAS} Collaboration, M.~Aaboud {\em et.~al.}, {\it {Search for heavy
  particles decaying into a top-quark pair in the fully hadronic final state in
  $pp$ collisions at $\sqrt{s} =13$ TeV with the ATLAS detector}},
  \href{http://arXiv.org/abs/1902.10077}{{\tt 1902.10077}}.

\bibitem{Aad:2015osa}
{\bf ATLAS} Collaboration, G.~Aad {\em et.~al.}, {\it {A search for high-mass
  resonances decaying to $\tau^{+}\tau^{-}$ in $pp$ collisions at $\sqrt{s}=8$
  TeV with the ATLAS detector}},  {\em JHEP} {\bf 07} (2015) 157
  [\href{http://arXiv.org/abs/1502.07177}{{\tt 1502.07177}}].

\bibitem{Aad:2019fac}
{\bf ATLAS} Collaboration, G.~Aad {\em et.~al.}, {\it {Search for high-mass
  dilepton resonances using 139 fb$^{-1}$ of $pp$ collision data collected at
  $\sqrt{s}=$13 TeV with the ATLAS detector}},
  \href{http://arXiv.org/abs/1903.06248}{{\tt 1903.06248}}.

\bibitem{atlasData}
{\bf ATLAS} Collaboration, G.~Aad {\em et.~al.}, {\it {Search for high-mass
  dilepton resonances using 139 fb$^{-1}$ of $pp$ collision data collected at
  $\sqrt{s}=$13 TeV with the ATLAS detector}},  2019.
\newblock https://www.hepdata.net/record/88425.

\bibitem{Degrande:2011ua}
C.~Degrande, C.~Duhr, B.~Fuks, D.~Grellscheid, O.~Mattelaer and T.~Reiter, {\it
  {UFO - The Universal FeynRules Output}},  {\em Comput. Phys. Commun.} {\bf
  183} (2012) 1201--1214 [\href{http://arXiv.org/abs/1108.2040}{{\tt
  1108.2040}}].

\bibitem{Alloul:2013bka}
A.~Alloul, N.~D. Christensen, C.~Degrande, C.~Duhr and B.~Fuks, {\it {FeynRules
  2.0 - A complete toolbox for tree-level phenomenology}},  {\em Comput. Phys.
  Commun.} {\bf 185} (2014) 2250--2300
  [\href{http://arXiv.org/abs/1310.1921}{{\tt 1310.1921}}].

\bibitem{Alwall:2014hca}
J.~Alwall, R.~Frederix, S.~Frixione, V.~Hirschi, F.~Maltoni, O.~Mattelaer,
  H.~S. Shao, T.~Stelzer, P.~Torrielli and M.~Zaro, {\it {The automated
  computation of tree-level and next-to-leading order differential cross
  sections, and their matching to parton shower simulations}},  {\em JHEP} {\bf
  07} (2014) 079 [\href{http://arXiv.org/abs/1405.0301}{{\tt 1405.0301}}].

\bibitem{Lim:2016wjo}
M.~Lim, F.~Maltoni, G.~Ridolfi and M.~Ubiali, {\it {Anatomy of double
  heavy-quark initiated processes}},  {\em JHEP} {\bf 09} (2016) 132
  [\href{http://arXiv.org/abs/1605.09411}{{\tt 1605.09411}}].

\bibitem{Chivukula:2017qsi}
R.~S. Chivukula, J.~Isaacson, K.~A. Mohan, D.~Sengupta and E.~H. Simmons, {\it
  {$R_K$ anomalies and simplified limits on $Z'$ models at the LHC}},  {\em
  Phys. Rev.} {\bf D96} (2017), no.~7 075012
  [\href{http://arXiv.org/abs/1706.06575}{{\tt 1706.06575}}].

\bibitem{Allanach:2017bta}
B.~C. Allanach, B.~Gripaios and T.~You, {\it {The case for future hadron
  colliders from $B \to K^{(*)} \mu^+ \mu^-$ decays}},  {\em JHEP} {\bf 03}
  (2018) 021 [\href{http://arXiv.org/abs/1710.06363}{{\tt 1710.06363}}].

\bibitem{Mangano:2018mur}
{\bf FCC} Collaboration, A.~Abada {\em et.~al.}, {\it {Future Circular Collider
  Study: Volume 1 - Physics Opportunities, Conceptual Design Report}},
  \href{http://arXiv.org/abs/http://inspirehep.net/record/1713706}{{\tt
  http://inspirehep.net/record/1713706}}.

\bibitem{Froggatt:1978nt}
C.~D. Froggatt and H.~B. Nielsen, {\it {Hierarchy of Quark Masses, Cabibbo
  Angles and CP Violation}},  {\em Nucl. Phys.} {\bf B147} (1979) 277--298.

\end{thebibliography}\endgroup

\end{document}